\documentclass[aip,jcp,reprint]{revtex4-1}
\usepackage{amsfonts}
\usepackage{amsmath}
\usepackage{amssymb}
\usepackage{graphicx}
\usepackage{bm}
\usepackage{hyperref}
\hypersetup{urlcolor=blue}
\numberwithin{lemma}{section}
\numberwithin{theorem}{section}
\numberwithin{remark}{section}
\numberwithin{corollary}{section}
\numberwithin{proposition}{section}
\numberwithin{definition}{section}
\numberwithin{example}{section}

\newcommand{\tran}{\ensuremath{^{\mathsf T}}}
\newcommand{\kb}{\ensuremath{k_\mathrm{B}}}
\newcommand{\ind}{\hspace*{2ex}}
\newcommand{\indd}{\hspace*{4ex}}
\newcommand{\inddd}{\hspace*{6ex}}

\usepackage{color}
\usepackage{soul}

\newcommand{\Rf}[1]{{\color{black}  #1}}
\newcommand{\our}[1]{{\color{black}  #1}}

\newcommand{\beginsupplement}{%
    \setcounter{table}{0}
    \renewcommand{\thetable}{S\arabic{table}}%
    \setcounter{figure}{0}
    \renewcommand{\thefigure}{S\arabic{figure}}%
    \setcounter{section}{0}
    \renewcommand{\thesection}{S\arabic{section}}%
    \setcounter{equation}{0}
    \renewcommand{\theequation}{S\arabic{equation}}%
    \setcounter{page}{1}
}

\begin{document}

\title{Geometric integrator for Langevin systems with quaternion-based rotational degrees of freedom and hydrodynamic interactions}

\author{R.~L. Davidchack}
\email{r.davidchack@le.ac.uk}
\affiliation{Department of Mathematics, University of Leicester, Leicester, LE1~7RH, UK}

\author{T.~E. Ouldridge}
\email{t.ouldridge@imperial.ac.uk}
\affiliation{Department of Bioengineering, Imperial College London, London SW7~2AZ, UK}

\author{M.~V. Tretyakov}
\email{Michael.Tretyakov@nottingham.ac.uk}
\affiliation{School of Mathematical Sciences, University of Nottingham, Nottingham, NG7~2RD, UK}

\begin{abstract}
We introduce new Langevin-type equations describing the rotational and translational motion of rigid bodies interacting through conservative and non-conservative forces, and hydrodynamic coupling. In the absence of non-conservative forces the Langevin-type equations sample from the canonical ensemble. The rotational degrees of freedom are described using quaternions, the lengths of which are exactly preserved by the stochastic dynamics. For the proposed Langevin-type equations, we construct a weak 2nd order geometric integrator which preserves the main geometric features of the continuous dynamics. 
{The integrator uses Verlet-type splitting for the deterministic part of Langevin equations appropriately combined with an exactly integrated Ornstein-Uhlenbeck process}. Numerical experiments are presented to illustrate both the new Langevin model and the numerical method for it, 
{as well as to demonstrate how inertia and the coupling of rotational and translational motion can introduce qualitatively distinct behaviours.}



\end{abstract}

\maketitle

\section{Introduction}
When modelling colloidal suspensions or solvated macromolecules, it is often convenient to treat the solvent molecules implicitly, thereby reducing the computational requirements of simulation\cite{SnookBook,Ouldridge_review_2017}. When doing so, an effective potential energy (in reality a free energy) is constructed between the remaining solute degrees of freedom \cite{Ouldridge_review_2017}. This effective potential, incorporating both direct and solvent-mediated contributions to the system's energy, can in principle reproduce the equilibrium distribution of the solute molecules having marginalised over the solvent configuration.

A given effective potential specifies the system's equilibrium distribution, but not its dynamics. Multiple dynamical models will in fact reach the same equilibrium distribution (see e.g. Refs.~\onlinecite{Ouldridge_review_2017,FrenkelBook,Milstein07} and references therein).  If dynamical properties are of interest, therefore, it is necessary to consider which of the possible dynamical models is most reasonable. One approach -- Langevin dynamics -- is to calculate generalised forces from the effective potential, and augment the resultant differential equations for the solute motion with additional noise and drag forces \cite{SnookBook, VanKampen2007}. These added forces model collisions with the implicit solvent and lead to diffusive motion of the solute particles.

A common approximation is to assume that the random and drag forces acting on each solute particle are independent. However, it is well-known that moving through a fluid sets up long-range flow fields that influence the drag experienced by other solute particles \cite{SnookBook}. These hydrodynamic interactions (HI) are often of fundamental importance to system properties. Famously, the presence of HI alters the scaling of polymer diffusion coefficients with length \cite{Zimm1956}. HI are known to strongly influence the rheological\cite{Park2016} and sedimentation\cite{Batchelor1971} properties of colloid suspensions, and are necessary to account for the diffusive behaviour of proteins within the cell \cite{Dlugosz2011,Skolnick2016}. Hydrodynamic coupling is also central to the motion and interaction of microscopic swimmers \cite{Najafi2004,Pooley2007}.

In the low Reynolds number limit, the hydrodynamic force experienced by a single solute particle is linear in the velocities of all particles \cite{SnookBook}. Hydrodynamic interactions in the Langevin formalism can then be encoded through a {\em friction matrix} \cite{SnookBook}, which relates the force experienced to the particle velocities. The friction matrix, and its inverse the mobility matrix, are in general dense and depend on the configuration of the system in a complex manner. Many approximations exist for calculating these matrices given a configuration (extensively reviewed in Ref. \onlinecite{SnookBook}), including methods for both long-range behaviour and so-called ``lubrication theory" which applies at short distances.

Given the functional form of the friction/mobility matrix, along with the effective potential, stochastic integrators can be constructed to implement the resultant ``Stokesian" dynamics. Ermak and McCammon's original first-order Euler-type integrator \cite{Ermak1978} is still widely used \cite{Dlugosz2011,Skolnick2016,Park2016}. This scheme assumes an over-damped limit, eliminating the particle momenta and directly updating particle positions based on the forces and mobility matrix. The method was subsequently generalised to incorporate rotational motion of solute particles \cite{Dickinson1985,Brady1988}. However, implementing rotational motion using, for example, Euler angles, can be problematic due to singularities in the equations of motion \cite{Evans77a}.

Unit quaternions, 4-dimensional unit vectors that can represent rigid-body orientation in 3D, are an alternative to Euler angles that avoid these singularities \cite{Evans77a}. Recently, several quaternion-based integrators have been demonstrated for Langevin dynamics in the absence of HI, both in the over-damped limit\cite{Davidchack09,Ilie2015,Davidchack15jcp} and beyond \cite{Davidchack09,Davidchack15jcp}. To date, however, little has been done to incorporate HI into quaternion-based integrators (a scheme is proposed in Ref.~\onlinecite{Ilie2015}, but not tested).

In this article we derive quaternion-based Langevin equations for the Stokesian dynamics of rigid bodies, based on the  Hamiltonian description of rigid-body dynamics in Ref.~\onlinecite{Miller02} and its extension to Langevin dynamics without HI in Refs.~\onlinecite{Davidchack09,Davidchack15jcp}.
 We then derive and demonstrate performance of a weak second-order geometric integrator building on the method of Ref.~\onlinecite{Davidchack15jcp}.
{
To do so, we concatenate the Verlet-type deterministic method for Hamiltonian dynamics of Ref.~\onlinecite{Miller02} with an exact integration of the Ornstein-Uhlenbeck process that combines HI and thermal noise.} Our approach does not assume an over-damped limit and naturally preserves the quaternion unit length up to machine precision.

{Our main contribution is to present and demonstrate this method for incorporating HI into quaternion-based integrators for rigid body dynamics. The specific approach, which does not assume an over-damped limit,  may be particularly useful in the following two contexts. First,  one can study systems with HI in non-over-damped regimes, where it is known\cite{Ai2017} that behaviour can  significantly differ from the over-damped case. Second, the proposed method works well even when the damping parameter is very large and hence, using it, one can get a good approximation of the over-damped (Brownian) dynamics. This feature is important since the direct  use of Brownian dynamics is often hindered by inefficiency of numerical integrators in handling stiff potentials \cite{Davidchack09,Davidchack15jcp}, and Langevin equations with large damping together with an appropriately constructed second-order geometric integrator can be more computationally efficient than simulating Brownian dynamics directly.}

The rest of the paper is organised as follows. After introducing  various quantities important for the problem setting in Section~\ref{sec:old}, we present new Langevin-type equations for rigid body dynamics with HI in Section~\ref{sec:new}. In Section~\ref{sec:met}, we derive a geometric integrator for the proposed Langevin-type equations, with the implementation details described in \Rf{Supplementary Material}. In Section~\ref{sec:num} we report results of a number of numerical experiments which test the constructed numerical method and also illustrate behaviour of the new Langevin model.

\section{Preliminaries
\label{sec:old}}
Consider a system of $n$ rigid bodies with the centre-of-mass coordinates $\mathbf{r}=(r^{1\,\mathsf{T}},\ldots ,r^{n\,\mathsf{T}})^{\mathsf{T}}\in \mathbb{R}^{3n}$, $r^{i}=(r_{1}^{i},r_{2}^{i},r_{3}^{i})^{\mathsf{T}}\in \mathbb{R}^{3}$ and orientations given by the unit quaternions $\mathbf{q}=(q^{1\,\mathsf{T}},\ldots ,q^{n\,\mathsf{T}})^{\mathsf{T}}$, $q^i = (q_0^i,q_{1}^{i},q_{2}^{i},q_{3}^{i})^\mathsf{T} \in \mathbb{S}^3$ (i.e., $|q^i| = 1$), immersed in an incompressible Newtonian fluid with viscosity $\eta$. 
{The unit quaternions, having three degrees of freedom each, are sufficient to encode the orientation of a rigid body.}  If the  interaction between particles is specified by an effective potential energy function $U(\mathbf{r},\mathbf{q})$, we can write a Hamiltonian for the $n$ rigid bodies in the form (see Ref.~\onlinecite{Miller02}):
\begin{equation}\label{a1}
H(\mathbf{r},\mathbf{p},\mathbf{q},\bm{\pi})=\sum_{i=1}^{n}\frac{%
p^{i\,\mathsf{T}}p^{i\,}}{2m_{i}}+\sum_{i=1}^{n}\sum_{l=1}^{3}\frac{1}{%
I_{l}^{i}}V_{l}(q^{i},\pi ^{i})+U(\mathbf{r},\mathbf{q}),
\end{equation}%
where $\mathbf{p}=(p^{1\,\mathsf{T}},\ldots ,p^{n\,\mathsf{T}})^{\mathsf{T}%
}\in \mathbb{R}^{3n}$, $p^{i}=(p_{1}^{i},p_{2}^{i},p_{3}^{i})^{\mathsf{T}%
}\in \mathbb{R}^{3},$ are the center-of-mass momenta conjugate to $\mathbf{r}$; $\bm{\pi}=(\pi ^{1\,\mathsf{T}},\ldots ,\pi ^{n\,\mathsf{T}})^{\mathsf{T}%
} $, $\pi ^{i}=(\pi _{0}^{i},\pi _{1}^{i},\pi _{2}^{i},\pi _{3}^{i})^{\mathsf{T}}$ are the angular momenta conjugate to $\mathbf{q}$ such that $q^{i\mathsf{T}}\pi ^{i}=0,$ i.e., $\pi ^{i}\in T^\ast_{q^{i}}\mathbb{S}^{3}$
which is the cotangent space to $\mathbb{S}^{3}$ at $q^{i}$.
The second term in (\ref{a1}) represents the rotational kinetic energy of the system with
\begin{equation} \label{a2}
V_{l}(q,\pi )=\frac{1}{8}\left[ \pi ^{\mathsf{T}}S_{l}q\right] ^{2},\ \
l=1,2,3,
\end{equation}
where the three constant $4$-by-$4$ matrices $S_{l}$ are
\begin{eqnarray*}
S_{1} &=&\left[
\begin{array}{cccc}
0 & -1 & 0 & 0 \\
1 & 0 & 0 & 0 \\
0 & 0 & 0 & 1 \\
0 & 0 & -1 & 0%
\end{array}%
\right] ,\ S_{2}=\left[
\begin{array}{cccc}
0 & 0 & -1 & 0 \\
0 & 0 & 0 & -1 \\
1 & 0 & 0 & 0 \\
0 & 1 & 0 & 0%
\end{array}%
\right] \\
S_{3} &=&\left[
\begin{array}{cccc}
0 & 0 & 0 & -1 \\
0 & 0 & 1 & 0 \\
0 & -1 & 0 & 0 \\
1 & 0 & 0 & 0%
\end{array}%
\right] ,
\end{eqnarray*}%
and $I_{l}^{i}$ are the principal moments of inertia of the rigid particle. 
{Note that $V_l(q^i,\pi^i)/I^i_l = I^i_l {\omega^i_l}^2/2$, where  $\omega^i_l$ (with $l = 1, 2, 3$)  are components of the angular velocity of particle $i$ in the body-fixed coordinate system.} We also introduce a diagonal matrix $\hat{D}^{i}=\mbox{diag}(1/I_{1}^{i},1/I_{2}^{i},1/I_{3}^{i})$ and a $4$-by-$3$ matrix
\begin{equation}\label{eq:hatS}
\hat{S}(q)=[S_{1}q,S_{2}q,S_{3}q]=\left[\begin{array}{ccc}
-q_{1} & -q_{2} & -q_{3} \\
 q_{0} & -q_{3} &  q_{2} \\
 q_{3} &  q_{0} & -q_{1} \\
-q_{2} &  q_{1} &  q_{0}
\end{array}\right].
\end{equation}
Note that $q^\mathsf{T}\hat{S}(q)=(0,0,0)$ and $\hat{S}^\mathsf{T}(q)\hat{S}(q)=\bm{1}_3$.

  The Newtonian equations of motion of the system described by the Hamiltonian in Eq.~(\ref{a1}), neglecting the influence of the solvent, are given by
\begin{eqnarray}\label{eq:hameom}
\frac{dr^i}{dt} &=& \frac{p^i}{m^i},\\
\frac{dp^i}{dt} &=& f^i(\mathbf{r},\mathbf{q}), \notag \\
\frac{dq^i}{dt} &=& \frac{1}{4}\hat{S}(q^i)\hat{D}^i\hat{S}^\mathsf{T}(q^i)\pi^i, \notag \\
\frac{d\pi^i}{dt} &=& \frac{1}{4}\hat{S}(\pi^i)\hat{D}^i\hat{S}^\mathsf{T}(q^i)\pi^i + F^i(\mathbf{r},\mathbf{q}), \notag
\end{eqnarray}
$i = 1, \ldots, n$, where $f^i(\mathbf{r},\mathbf{q}) = -\nabla_{r^i} U(\mathbf{r},\mathbf{q}) \in \mathbb{R}^3$ is the translational force acting on particle $i$ and $F^i(\mathbf{r},\mathbf{q}) = -\tilde{\nabla}_{q^i} U(\mathbf{r},\mathbf{q})$ is the rotational force.

{It is important} that, while $\nabla_{r}$ is the gradient in the Cartesian coordinates in $\mathbb{R}^{3}$, $\tilde{\nabla}_{q}$ is the directional derivative tangent to the three-dimensional sphere $\mathbb{S}^{3}$, implying that ${q}^{i\mathsf{T}} F^i = 0$.  The directional derivative can be expressed in terms of the four-dimensional gradient as $\tilde{\nabla}_q = (\bm{1}_4 - q q^\mathsf{T})\nabla_q$, where $\bm{1}_4$ is the four-dimensional identity matrix.  The rotational force can also be calculated as $F^i = 2\hat{S}(q^i)\tau^i$, where $\tau^i \in \mathbb{R}^3$ is the torque on molecule $i$ in the {\em body-fixed coordinate frame} (i.e. with axes aligned with the principal axes of the rigid body and rotating with it).

In addition to the interaction forces, the particles also experience drag forces due to their motion relative to the fluid and stochastic forces due to thermal fluctuations.  In the low Reynolds number regime, the hydrodynamic force $f_\mathrm{h}^i$ and torque $T_\mathrm{h}^i$ experienced by particle $i$ depend linearly on the linear and angular velocities, $v^i = p^i/m^i$ and $\Omega^i$, through a $6n$-by-$6n$ coordinate-dependent {\em friction matrix}
\begin{equation}\label{eq:friction}
\xi(\mathbf{r},\mathbf{q})=\left[\begin{array}{cc}
  ^{\mathrm{tt}}\xi (\mathbf{r},\mathbf{q}) & ^{\mathrm{tr}}\xi (\mathbf{r},\mathbf{q}) \\
  ^{\mathrm{rt}}\xi (\mathbf{r},\mathbf{q}) & ^{\mathrm{rr}}\xi (\mathbf{r},\mathbf{q})
\end{array}\right],
\end{equation}
as follows
\begin{eqnarray}\label{eq:hydroforce}
  f^i_\mathrm{h} &=& -\sum_{j=1}^n\left(^\mathrm{tt}\xi^{(i,j)}(\mathbf{r},\mathbf{q})\,v^j +\,^\mathrm{tr}\xi^{(i,j)}(\mathbf{r},\mathbf{q})\,\Omega^j \right),\\
  T^i_\mathrm{h} &=& -\sum_{j=1}^n\left(^\mathrm{rt}\xi^{(i,j)}(\mathbf{r},\mathbf{q})\,v^j +\,^\mathrm{rr}\xi^{(i,j)}(\mathbf{r},\mathbf{q})\,\Omega^j \right),\nonumber\\
&& \,\,\, i = 1, \ldots, n, \nonumber
\end{eqnarray}
where $T^i_\mathrm{h}$ and $\Omega^i$ are torques and angular velocities in the {\em space-fixed coordinate frame}.  The left superscripts t and r denote components of the friction matrix $\xi$ coupling the translational and rotational degrees of freedom, respectively.  Each sub-matrix $^{ab}\xi(\mathbf{r},\mathbf{q})$ in Eq.~(\ref{eq:friction}) contains $n^2$ $3$-by-$3$ blocks $^{ab}\xi^{(i,j)}(\mathbf{r},\mathbf{q})$, $i,j = 1, \ldots, n$, $a, b = $~t,~r.  Matrix $\xi(\mathbf{r},\mathbf{q})$ is symmetric, so that $^\mathrm{tt}\xi^{(i,j)} = \,^\mathrm{tt}\xi^{(j,i)\,\mathsf{T}}$, $^\mathrm{tr}\xi^{(i,j)} = \,^\mathrm{rt}\xi^{(j,i)\,\mathsf{T}}$, and $^\mathrm{rr}\xi^{(i,j)} = \,^\mathrm{rr}\xi^{(j,i)\,\mathsf{T}}$. The calculation of the friction matrix is, in general, a complex problem for which many approximate results exist \cite{SnookBook}. The best choice of a friction matrix for a given problem is beyond the scope of this paper -- we simply present a method into which any well-defined positive-definite symmetric friction matrix $\xi$ can be substituted. We note that in HI calculations one usually uses spherical particles as an approximation due to the difficulty in calculating hydrodynamic coupling for non-spherical particles. In the case of this approximation the friction matrix $\xi$ depends only on the centre-of-mass positions $\mathbf{r}$. By including dependence of $\xi$ on the orientation of particles expressed via quaternions $\mathbf{q}$, we allow for hydrodynamic interactions between non-spherical objects. One approach would be to model non-spherical particles as rigid clusters of spherical particles \cite{Kutteh10} for the purposes of calculating $\xi$, in which case $\mathbf{r}$ and $\mathbf{q}$ would represent positions and orientations of the clusters.

The angular velocity $\Omega^i$ in the space-fixed coordinate frame is related to the conjugate momentum $\pi^i$ as follows
\begin{equation}\label{eq:angular}
\Omega^i = A^\mathsf{T}(q^i) \omega^i = \tfrac{1}{2}A^\mathsf{T}(q^i)\hat{D}^i\hat{S}^\mathsf{T}(q^i)\pi^i,
\end{equation}
where $\omega^i$ is the angular velocity in the {\em body-fixed coordinate frame} (with coordinate axes aligned with the principal directions of the rigid body), and the rotation matrix,
\begin{equation}\label{eq:rotmat}
A(q)=2\left[ \begin{array}{ccc}
q_{0}^{2}+q_{1}^{2}-\frac{1}{2} & q_{1}q_{2}+q_{0}q_{3} & q_{1}q_{3}-q_{0}q_{2} \\
q_{1}q_{2}-q_{0}q_{3} & q_{0}^{2}+q_{2}^{2}-\frac{1}{2} & q_{2}q_{3}+q_{0}q_{1} \\
q_{1}q_{3}+q_{0}q_{2} & q_{2}q_{3}-q_{0}q_{1} & q_{0}^{2}+q_{3}^{2}-\frac{1}{2}
\end{array}\right],
\end{equation}
transforms from space-fixed to body-fixed frame, while its transpose $A^\mathsf{T}(q)$ transforms from body-fixed to space-fixed frame.  For example, the torque $\tau^i$ on particle $i$ in the body-fixed frame is related to the torque in the space-fixed frame $T^i$ by $\tau^i = A(q^i)T^i$.

The stochastic forces can be modelled by white noise \our{so that} in the absence of any other external forces, the equilibrium probability distribution of the system is Gibbsian with temperature $T$: $\rho (\mathbf{r},\mathbf{p},\mathbf{q},\bm{\pi})\varpropto \exp (-\beta H(\mathbf{r},\mathbf{p},\mathbf{q},\bm{\pi }))$, where $\beta^{-1} = k_B T$.

\section{New Langevin-type equations for rigid body dynamics with hydrodynamic interactions
\label{sec:new}}
Based on Section~\ref{sec:old}, 
{we propose the following Langevin-type equations (in the form of It\^o) to model the}  time evolution of rigid bodies under influence of conservative forces, hydrodynamic interactions and thermal noise:
\begin{eqnarray}\label{la1}
dR^{i} &=&\frac{P^{i}}{m^{i}}dt,\ \ R^{i}(0)=r^{i},   \\
dP^{i} &=&f^{i}(\mathbf{R},\mathbf{Q})dt-\sum_{j=1}^{n}
\,^{\mathrm{tt}}\xi^{(i,j)}(\mathbf{R},\mathbf{Q})\frac{P^{j}}{m^j}dt \notag \\
&&-\frac{1}{2}\sum_{j=1}^{n}\,^{\mathrm{tr}}\xi^{(i,j)}(\mathbf{R},\mathbf{Q})
A^\mathsf{T}(Q^{j})\hat{D}^{j}\hat{S}^\mathsf{T}(Q^{j})\Pi ^{j}dt \notag \\
&&+\!\sum_{j=1}^{n}\ ^{\mathrm{tt}}b^{(i,j)}(\mathbf{R},\mathbf{Q})dw^{j}(t)+%
\sum_{j=1}^{n}\ ^{\mathrm{tr}}b^{(i,j)}(\mathbf{R},\mathbf{Q})dW^{j}(t),\notag \\
&&P^{i}(0)=p^{i}, \notag
\end{eqnarray}
\begin{eqnarray} \label{la2}
dQ^{i} &=&\frac{1}{4}\hat{S}(Q^{i})\hat{D}^{i}\hat{S}^\mathsf{T}(Q^{i})
\Pi^{i}dt, \\
&& Q^{i}(0)=q^{i},\ \ |q^{i}|=1, \notag \\
d\Pi^{i} &=&\frac{1}{4}\hat{S}(\Pi^{i})\hat{D}^{i}\hat{S}^\mathsf{T}(Q^{i})\Pi^{i}dt
+{F^{i}}(\mathbf{R},\mathbf{Q})dt  \notag \\
&&-\sum_{j=1}^{n}\check{S}(Q^{i})\,^{\mathrm{rr}}\xi^{(i,j)}(\mathbf{R},\mathbf{Q})
A^\mathsf{T}(Q^{j})\hat{D}^{j}\hat{S}^\mathsf{T}(Q^{j})\Pi^{j}dt \notag \\
&&-2\sum_{j=1}^{n}\check{S}(Q^{i})
\,^{\mathrm{rt}}\xi^{(i,j)}(\mathbf{R},\mathbf{Q})\frac{P^{j}}{m^j}dt  \notag \\
&&+2\sum_{j=1}^{n}\check{S}(Q^{i})\,^{\mathrm{rr}}b^{(i,j)}(\mathbf{R},\mathbf{Q})dW^{j}(t) \notag \\
&&+2\sum_{j=1}^{n}\check{S}(Q^{i})\ ^{\mathrm{rt}}b^{(i,j)}(\mathbf{R},\mathbf{Q})dw^{j}(t),\notag \\
&& \Pi^{i}(0)=\pi ^{i},\ \ q^{i\mathsf{T}}\pi ^{i}=0,\ \ i=1,\ldots ,n,   \notag
\end{eqnarray}
where $^{\mathrm{tt}}b^{(i,j)}(\mathbf{r},\mathbf{q}),$ $^{\mathrm{tr}}b^{(i,j)}(\mathbf{r},\mathbf{q}),$ $^{\mathrm{rr}}b^{(i,j)}(\mathbf{r},\mathbf{q}),$ and $^{\mathrm{rt}}b^{(i,j)}(\mathbf{r},\mathbf{q})$, are $3\times 3$-matrices; $(\mathbf{w}^\mathsf{T},\mathbf{W}^\mathsf{T})^\mathsf{T}=(w^{1\mathsf{T}},\ldots ,w^{n\mathsf{T}},W^{1\mathsf{T}},\ldots,W^{n\mathsf{T}})^\mathsf{T}$ is a $(3n+3n)$-dimensional standard Wiener process with $w^{i}=(w_{1}^{i},w_{2}^{i},w_{3}^{i})^\mathsf{T}$ and $W^{i}=(W_{1}^{i},W_{2}^{i},W_{3}^{i})^\mathsf{T}$. We also define
\begin{equation}\label{eq:checkS}
\check{S}(q)=\hat{S}(q)A(q)=\left[
\begin{array}{ccc}
-q_{1} & -q_{2} & -q_{3} \\
q_{0} & q_{3} & -q_{2} \\
-q_{3} & q_{0} & q_{1} \\
q_{2} & -q_{1} & q_{0}%
\end{array}%
\right]
\end{equation}
given that $|q| = 1$. Note that $q^\mathsf{T}\check{S}(q)=(0,0,0)$, $\check{S}^\mathsf{T}(q)\check{S}(q)=\bm{1}_3$, and $A(q)=\hat{S}^\mathsf{T}(q)\check{S}(q)$.

The Langevin model (\ref{la1})-(\ref{la2}) has the  following \Rf{important} properties.
(i) The solution of (\ref{la1})-(\ref{la2}) preserves the quaternion lengths:
\begin{equation}\label{a211}
|Q^{i}(t)|=1,\ \ i=1,\ldots ,n\,,\ \ \mbox{for all~}t\geq 0,
\end{equation}
\Rf{since $dQ^{i}$ is orthogonal to $Q^{i}$ by the properties of the matrix $\hat{S}(Q^i)$.}

(ii) The solution of (\ref{la1})-(\ref{la2}) preserves orthogonality of $\mathbf{Q}(t)$ and $\mathbf{\Pi }(t)$:
\begin{equation}\label{qpi0}
\mathbf{Q}^{\mathsf{T}}(t)\mathbf{\Pi }(t)=0,\ \ \mbox{for all~}t\geq 0,
\end{equation}
\Rf{since $d\Pi^{i}$ is orthogonal to $Q^{i}$  and $\Pi^{i}$ is orthogonal to $dQ^{i}$, which can be shown using the properties of the $\hat{S}(Q^i)$ and $\check{S}(Q^i)$  matrices, and the torque $F^i$.}

(iii) The It\^{o} interpretation of the system of SDEs (\ref{la1})-(\ref{la2}) coincides with its Stratonovich interpretation, \Rf{since noise terms depend on position and orientation, but act directly on the generalised momenta only}.

(iv) Assume that the solution $X(t)=(\mathbf{R}^{\mathsf{T}}(t)$, $\mathbf{P}^{\mathsf{T}}(t),\mathbf{Q}^{\mathsf{T}}(t),\mathbf{\Pi}^{\mathsf{T}}(t))^{\mathsf{T}}$ of (\ref{la1})-(\ref{la2}) is an ergodic process \cite{Has,Soize} on
\begin{eqnarray*}
\ \ \mathbb{D} &=&\{x=(\mathbf{r}^{\mathsf{T}},\mathbf{p}^{\mathsf{T}},%
\mathbf{q}^{\mathsf{T}},\bm{\pi}^{\mathsf{T}})^{\mathsf{T}}\in \mathbb{R}%
^{14n}: \\
&&\ \ |q^{i}|=1,(q^{i})^{\top }\pi ^{i}=0,\ \ i=1,\ldots ,n\}.
\end{eqnarray*}
Then the invariant measure of $X(t)$ is Gibbsian with the density
$\rho (\mathbf{r},\mathbf{p},\mathbf{q},\bm{\pi})$,
\begin{equation}\label{eq:gibbs}
\rho (\mathbf{r},\mathbf{p},\mathbf{q},\bm{\pi})\varpropto \exp (-\beta H(\mathbf{r},\mathbf{p},\mathbf{q},\bm{\pi})),
\end{equation}
if the following condition holds
\begin{equation}\label{eq:bbxi}
  b(\mathbf{r},\mathbf{q}) b^\mathsf{T}(\mathbf{r},\mathbf{q}) = \frac{2}{\beta}\xi(\mathbf{r},\mathbf{q}),
\end{equation}
where
\begin{equation}\label{eq:bb}
b(\mathbf{r},\mathbf{q})=\left[\begin{array}{cc}
^{\mathrm{tt}}b (\mathbf{r},\mathbf{q}) & ^{\mathrm{tr}}b (\mathbf{r},\mathbf{q}) \\
^{\mathrm{rt}}b (\mathbf{r},\mathbf{q}) & ^{\mathrm{rr}}b (\mathbf{r},\mathbf{q})
\end{array}\right]
\end{equation}
and each sub-matrix $^{ab}b(\mathbf{r},\mathbf{q})$ contains $n^2$ $3$-by-$3$ blocks $^{ab}b^{(i,j)}(\mathbf{r},\mathbf{q})$, $i,j = 1, \ldots, n$, $a, b = $~t,~r.

\Rf{From the statistical physics viewpoint, the fluctuation-dissipation theorem embodied by the constraint (\ref{eq:bbxi}) ensures that noise and damping are perfectly balanced to produce the correct Gibbsian distribution in equilibrium. Given the non-trivial Langevin system (\ref{la1})-(\ref{la2}), the form of  this relation is not immediately obvious, but can be demonstrated by verifying that the stationary Fokker-Planck equation corresponding to (\ref{la1})-(\ref{la2}), (\ref{eq:bbxi}) is satisfied with $\rho$ given by Eq.~(\ref{eq:gibbs}) and $H$ by Eq.~(\ref{a1}).}


We note that if $^\mathrm{tr}\xi^{(i,j)} = \,^\mathrm{rt}\xi^{(j,i)}=0$, $^\mathrm{tr}b^{(i,j)} = \,^\mathrm{rt}b^{(j,i)}=0$, and $^\mathrm{tt}\xi^{(i,j)}$,  $^\mathrm{rr}\xi^{(i,j)}$, $^\mathrm{tt}b^{(i,j)}$,  and $^\mathrm{rr}b^{(i,j)}$ are appropriately chosen diagonal constant matrices, then the system (\ref{la1})-(\ref{la2}) degenerates to the Langevin thermostat for rigid bodies from Ref.~\onlinecite{Davidchack15jcp}.

The Langevin system (\ref{la1})-(\ref{la2}) is driven by the conservative forces, hydrodynamic interactions and thermal noise. It is worthwhile (see e.g.~Refs.~\onlinecite{Dunkel2009,ReichertDiss,Joubad2015} and the example in Section~\ref{sec:ne3} here) to generalize this system by including non-conservative, possibly time-dependent, forces and torques $\tilde{\mathbf{f}}(t,\mathbf{r},\mathbf{q})$ and $\tilde{\mathbf{F}}(t,\mathbf{r},\mathbf{q})$, i.e., to consider the system of Langevin-type equations 
{of the form (\ref{la1})-(\ref{la2}) with the additional terms
$\tilde{\mathbf{f}^i}(t,\mathbf{R},\mathbf{Q})dt$ in the equations for $P^i$ and
$\tilde{\mathbf{F}^i}(t,\mathbf{R},\mathbf{Q})dt$ in the equations for $\Pi^{i}$.}
The first three properties stated above for the model (\ref{la1})-(\ref{la2}) are also true for the system with non-conservative forces.

\section{Numerical integrator
\label{sec:met}}
In this section we propose a weak 2nd order numerical integrator for (\ref{la1})-(\ref{la2}), which is a generalization of the Langevin C method from Ref.~\onlinecite{Davidchack15jcp}. 
{To obtain the integrator, we (similarly to Ref.~\onlinecite{Davidchack15jcp}) exploit the efficient numerical scheme for rigid bodies' Hamiltonian dynamics from Ref.~\onlinecite{Miller02}, which is based on the Verlet-type splitting, and use exact integration of the Ornstein-Uhlenbeck process that combines HI and thermal noise as can be seen below. Thus,} the new integrator is based on splitting (\ref{la1})-(\ref{la2}) into the deterministic Hamiltonian system (\ref{eq:hameom}) and the Ornstein-Uhlenbeck-type SDEs for \Rf{hydrodynamically-coupled diffusion}:
\begin{equation}\label{ou4}
dY=-\tilde{\xi}(\mathbf{r},\mathbf{q})Y+\tilde{b}(\mathbf{r},\mathbf{q})d\tilde{\mathbf{W}}(t),
\end{equation}%
where
\begin{eqnarray*}
\tilde{\xi}(\mathbf{r},\mathbf{q}) &:=& \left[\begin{array}{cc}
^{\mathrm{tt}}\tilde{\xi}(\mathbf{r},\mathbf{q}) & ^{\mathrm{tr}}\tilde{\xi}(\mathbf{r},\mathbf{q}) \\
^{\mathrm{rt}}\tilde{\xi}(\mathbf{r},\mathbf{q}) & ^{\mathrm{rr}}\tilde{\xi}(\mathbf{r},\mathbf{q})
\end{array}\right],\\
\tilde{b}(\mathbf{r},\mathbf{q})&:=& \left[\begin{array}{cc}
^{\mathrm{tt}}\tilde{b}(\mathbf{r},\mathbf{q}) & ^{\mathrm{tr}}\tilde{b}(\mathbf{r},\mathbf{q}) \\
^{\mathrm{rt}}\tilde{b}(\mathbf{r},\mathbf{q}) & ^{\mathrm{rr}}\tilde{b}(\mathbf{r},\mathbf{q})%
\end{array}%
\right],\\
Y &:=& \left[\begin{array}{c} \mathbf{\tilde P}\\ \bm{\tilde \Pi} \end{array}\right],\ \
\tilde{\mathbf{W}}(t) := \left[\begin{array}{c} \mathbf{w}(t) \\\mathbf{W}(t) \end{array}\right],
\end{eqnarray*}%
and
\begin{eqnarray*} \label{ou2}
^{\mathrm{tt}}\tilde{\xi}^{(i,j)}(\mathbf{r},\mathbf{q}) &:=&
\frac{1}{m^{j}}\,^{\mathrm{tt}}\xi^{(i,j)}(\mathbf{r},\mathbf{q}),\\
^{\mathrm{tr}}\tilde{\xi}^{(i,j)}(\mathbf{r},\mathbf{q}) &:=&
\frac{1}{2}\,^{\mathrm{tr}}\xi^{(i,j)}(\mathbf{r},\mathbf{q})A^\mathsf{T}(q^{j})\hat{D}^{j}\hat{S}^\mathsf{T}(q^{j}),\notag \\
^{\mathrm{rt}}\tilde{\xi}^{(i,j)}(\mathbf{r},\mathbf{q}) &:=&
\frac{2}{m^{j}}\check{S}(q^{i})\,^{\mathrm{rt}}\xi ^{(i,j)}(\mathbf{r},\mathbf{q}),  \notag \\
^{\mathrm{rr}}\tilde{\xi}^{(i,j)}(\mathbf{r},\mathbf{q}) &:=&
\check{S}(q^{i})\,^{\mathrm{rr}}\xi^{(i,j)}(\mathbf{r},\mathbf{q})A^\mathsf{T}(q^{j})\hat{D}^{j}\hat{S}^\mathsf{T}(q^{j}), \notag \\
^{\mathrm{rr}}\tilde{b}^{(i,j)}(\mathbf{r},\mathbf{q}) &:=& 2\check{S}(q^{i})\,^{\mathrm{rr}}b^{(i,j)}(\mathbf{r},\mathbf{q}), \notag \\
^{\mathrm{rt}}\tilde{b}^{(i,j)}(\mathbf{r},\mathbf{q}) &:=& 2\check{S}(q^{i})\,^{\mathrm{rt}}b^{(i,j)}(\mathbf{r},\mathbf{q}),  \notag \\
^{\mathrm{tt}}\tilde{b}^{(i,j)}(\mathbf{r},\mathbf{q}) &:=&
^{\mathrm{tt}}b^{(i,j)}(\mathbf{r},\mathbf{q}),\notag \\
^{\mathrm{rr}}\tilde{b}^{(i,j)}(\mathbf{r},\mathbf{q}) &:=&
^{\mathrm{rr}}b^{(i,j)}(\mathbf{r},\mathbf{q}).\notag
\end{eqnarray*}
In (\ref{ou4}), $\mathbf{r}$ and $\mathbf{q}$ are fixed.  Note that, unlike $\xi (\mathbf{r},\mathbf{q})$, the matrix $\tilde{\xi}(\mathbf{r},\mathbf{q})$ is not symmetric.

The solution of the linear SDEs with additive noise (\ref{ou4}) is given by
\begin{equation}\label{ou5}
Y(t)=\mathrm{e}^{-\tilde{\xi}(\mathbf{r},\mathbf{q})t}Y(0)+\int_{0}^{t}\mathrm{e}^{-\tilde{\xi}(\mathbf{r},\mathbf{q})(t-s)}\tilde{b}(\mathbf{r},\mathbf{q})d\tilde{\mathbf{W}}(s).
\end{equation}
The $7n$-dimensional vector $\int_{0}^{t}\mathrm{e}^{-\tilde{\xi}(\mathbf{r},\mathbf{q})(t-s)}\tilde{b}(\mathbf{r},\mathbf{q})d\tilde{\mathbf{W}}(t)$ is Gaussian with zero mean and covariance
\begin{equation}\label{ou6}
C(t;\mathbf{r},\mathbf{q})=\int_{0}^{t}\mathrm{e}^{-\tilde{\xi}(\mathbf{r},\mathbf{q})(t-s)}%
\tilde{b}(\mathbf{r},\mathbf{q})\tilde{b}^{\mathsf{T}}(\mathbf{r},\mathbf{q})\mathrm{e}^{-%
\tilde{\xi}^{\mathsf{T}}(\mathbf{r},\mathbf{q})(t-s)}ds.
\end{equation}
Introducing a $7n$-by-$6n$ matrix $\sigma (t;\mathbf{r},\mathbf{q})$ such that
\begin{equation}\label{lbec}
\sigma (t;\mathbf{r},\mathbf{q})\sigma ^{\mathsf{T}}(t;\mathbf{r},\mathbf{q})=C(t;\mathbf{r},\mathbf{q}),
\end{equation}
we can write Eq.~(\ref{ou5}) in the form
\begin{equation}\label{ou7}
Y(t)=\mathrm{e}^{-\tilde{\xi}(\mathbf{r},\mathbf{q})t}Y(0)+\sigma(t;\mathbf{r},\mathbf{q})\chi,
\end{equation}%
where $\chi$ is a $6n$-dimensional vector consisting of independent Gaussian random variables with zero mean and unit variance.  Details of the evaluation of the covariance integral (\ref{ou6}) and matrix $\sigma(t;\mathbf{r},\mathbf{q})$ can be found in Appendix~\ref{app:a}.

{We now present the integrator itself. In each time step of size $h$, we perform half a step of the Verlet-type integrator for Hamiltonian dynamics,\cite{Miller02} followed by a full time step of the Ornstein-Uhlenbeck process in (\ref{ou7}), and finally a second half step of the Verlet-type integrator. }
Starting from the initial conditions $\mathbf{P}_{0}=\mathbf{p}$, $\mathbf{R}_{0}=\mathbf{r}$, $\mathbf{Q}_{0}=\mathbf{q}$, $|q^{i}|=1$, $i=1,\ldots ,n$, $\bm{\Pi}_{0}=\bm{\pi}$, $\mathbf{q}^{\mathsf{T}}\bm{\pi}=0$, the numerical integrator for (\ref{la1})-(\ref{la2}) takes the form
\begin{eqnarray}\label{langC}
\mathcal{P}_{1,k}^{i} &=& {P}_{k}^{i}+\frac{h}{2}{f}^{i}(\mathbf{R}_{k},\mathbf{Q}_{k}), \\
\mathit{\Pi}_{1,k}^{i} &=& \Pi_{k}^{i}+\frac{h}{2}F{^{i}}(\mathbf{R}_{k},\mathbf{Q}_{k}), \notag \\
\mathcal{R}_{1,k}^{i} &=& {R}_{k}^{i}+\frac{h}{2}\frac{\mathcal{P}_{1,k}^{i}}{m^{i}}, \notag \\
(\mathcal{Q}_{1,k}^{i},&&\hspace{-2.5ex}\mathit{\Pi}_{2,k}^{i}) = \Psi^{-}_{h/2}(Q_{k}^{i},\mathit{\Pi}_{1,k}^{i}), \notag \\[1ex]
\left[\begin{array}{c} \mathcal{P}_{2,k} \\ \mathit{\Pi}_{3,k} \end{array}\right] &=&
\mathrm{e}^{-\tilde{\xi}(\mathcal{R}_{1,k},\mathcal{Q}_{1,k})h}
\left[\begin{array}{c} \mathcal{P}_{1,k} \\ \mathit{\Pi}_{2,k} \end{array}\right]
+ \sigma(h;\mathcal{R}_{1,k},\mathcal{Q}_{1,k})\chi_{k}, \notag \\
{R}_{k+1}^{i} &=& \mathcal{R}_{1,k}^{i} + \frac{h}{2}\frac{\mathcal{P}_{2,k}^{i}}{m^{i}}, \notag \\
(Q_{k+1}^{i},&&\hspace{-2.5ex}\mathit{\Pi}_{4,k}^{i}) =\Psi_{h/2}^{+}(\mathcal{Q}_{1,k}^{i},\mathit{\Pi}_{3,k}^{i}),\notag \\
P_{k+1}^{i} &=& \mathcal{P}_{2,k}^{i} + \frac{h}{2}{f}^{i}(\mathbf{R}_{k+1},\mathbf{Q}_{k+1}), \notag \\
\Pi_{k+1}^{i} &=&\mathit{\Pi}_{4,k}^{i}+\frac{h}{2}F^{i}(\mathbf{R}_{k+1},\mathbf{Q}_{k+1}),\notag
\end{eqnarray}
$i = 1,\ldots,n$, where $\chi_{k}$ is a $6n$-dimensional vector with components being i.i.d.~Gaussian random variables with zero mean and unit variance. We note in passing that for weak convergence it is sufficient\cite{MT1} to use the simpler law: $P(\theta =0)=2/3,\ \ P(\theta =\pm \sqrt{3})=1/6$  for the components of  $\chi_{k}$.

As in Ref.~\onlinecite{Davidchack15jcp} (see also Refs.~\onlinecite{Miller02,Davidchack09}), we use exact rotations around each principal axis written as the maps $\Psi_{t,l}(q,\pi):\ (q,\pi)\mapsto (\mathcal{Q},\mathit{\Pi})$ defined by:
\begin{eqnarray}\label{a24}
\mathcal{Q}  &=& \cos (\zeta_{l}t)q +  \sin (\zeta_{l}t)S_{l}q, \\
\mathit{\Pi} &=& \cos (\zeta_{l}t)\pi +\sin (\zeta_{l}t)S_{l}\pi, \notag
\end{eqnarray}
where $\zeta_{l}=\frac{1}{4I_{l}}\pi ^{\mathsf{T}}S_{l}q$. Based on (\ref{a24}), the composite maps $\Psi^{\pm}_{t,l}(q,\pi):\ (q,\pi)\mapsto (\mathcal{Q},\mathit{\Pi})$ used in (\ref{langC}) are defined as
\begin{eqnarray}
\Psi_{t}^{-}=\Psi _{t,3}\circ \Psi _{t,2}\circ \Psi _{t,1}, \label{a25}\\
\Psi_{t}^{+}=\Psi _{t,1}\circ \Psi _{t,2}\circ \Psi _{t,3} \nonumber,
\end{eqnarray}
where \textquotedblleft $\circ $\textquotedblright\ denotes function composition, i.e., $(g\circ f)(x)=g(f(x))$. Implementation details for the method (\ref{langC}) are given in
{Section S4 of the Supplementary Material.}

The proposed method (\ref{langC}) has the following key properties:
(i) it is quasi-symplectic, in the sense that it degenerates to Langevin C from Ref.~\onlinecite{Davidchack15jcp} \our{when (\ref{la1})-(\ref{la2}) degenerates to Langevin thermostat for rigid bodies (without HI) from
Ref.~\onlinecite{Davidchack15jcp}} (see also Refs.~\onlinecite{MT03,MT1});
(ii) it preserves $|Q_k^{j}|=1,\ j=1,\ldots, n,$ for all $t_k\geq 0$ automatically, \Rf{since $Q$ is only updated by exact rotations};
 (iii) it preserves $Q_k^{j\,\mathsf{T}}\Pi_k^{j}=0\,,\ j=1,\ldots ,n,$ for $t_k\geq 0$ automatically, \Rf{since $\Pi$ is updated through increments that are orthogonal to $Q$ or exact rotations of $Q,\Pi$ system};
(iv) only a single evaluation of forces and the friction matrix per step is required;
(v) it is of weak order $2$.

The weak convergence of the method is proved by the standard arguments as follows. \Rf{It is straightforward to apply the one-step approximation of a standard weak-second-order Taylor-type method (see Ref.~\onlinecite[p. 94]{MT1}) to  the evolution of $X(t)=(\mathbf{R}^{\mathsf{T}}(t)$, $\mathbf{P}^{\mathsf{T}}(t),\mathbf{Q}^{\mathsf{T}}(t),\mathbf{\Pi}^{\mathsf{T}}(t))^{\mathsf{T}}$ during the period of one step $h$ directly from the  SDEs (\ref{la1})-(\ref{la2}).
This approach is inappropriate for creating an efficient numerical scheme, in part because it requires computation of derivatives of forces and the friction matrix, and also because it fails to reflect the underlying constraints under which the system evolves (e.g.~$|Q_k^{j}|=1$). However, we can compare this one-step approximation to an expansion of the one-step approximation corresponding to our method  (\ref{langC}). By matching appropriate moments of the increments of the two one-step approximations up to the third order in $h$, we confirmed the weak 2nd order accuracy of our method (\ref{langC}) via the  general weak convergence theorem (see Ref.~\onlinecite[pp.100-101]{MT1}). The necessary algebra is conceptually simple, but tedious.
A detailed example of such a proof by comparison can be found, e.g. in Ref.~\onlinecite[Section 10]{MT97sinum}.}

Even though the proposed integrator preserves the constraints $|q^i| - 1 = 0$ and $q^{i\mathsf{T}} \pi^i = 0$, in practice these quantities gradually deviate from zero in the course of a long simulation due to the finite accuracy of double precision arithmetic.  In our simulations we observe that starting with a deviation of the order of $10^{-16}$ in the double-precision computations, the maximum deviation grows to about $10^{-12}$ at the end of the simulation run for the test simulation in \Rf{Section~S2 of the Supplementary Material}, independent of the time step $h$.  As we demonstrated in Ref.~\onlinecite{Davidchack15jcp}, the deviation of $q^{i\mathsf{T}} \pi^i$ from zero does not have any effect on physically relevant quantities.  On the other hand, the deviation of $|q^i|$ from 1 does have an effect on measured quantities.  Therefore, we recommend re-normalising quaternion coordinates, especially in very long simulations.  Since the computational cost of such normalisation is relatively insignificant, it can be done even after every step.

{At the same time, we emphasize that for the stability of a numerical method it is crucial to preserve key geometric features of the continuous dynamics in the discrete approximation. In particular, it is well known (see e.g. Refs.~\onlinecite{DLM97,HLW02,Johan10} and references therein) that if continuous dynamics live on a manifold (in this case $q^{i\mathsf{T}} \pi^i = 0$ and $|q^i|=1$),  then to ensure long time stability of a numerical method it should stay on the same manifold as well, which is the case for the numerical integrator developed in this paper. The renormalization of quaternions discussed in the previous paragraph is just dealing with the (relatively small) round-off error from machine precision, not with the numerical integration error for an integrator that does not naturally conserve quaternion length. We also note in passing the benefit of using the quaternion representation of rotations in comparison with rotational matrices in constructing numerical integrators. In the first case the deviation of $|q^i|$ from 1 due to round-off errors is easy to correct as described above; in the second case when the orthogonal matrices lose their orthogonality property due to round-off errors, it is not a trivial task to make the matrices orthogonal again.}

{We remark that} in Ref.~\onlinecite{Davidchack15jcp} we constructed three weak 2nd order geometric integrators (Langevin A, B and C) for  a Langevin thermostat (without HI). The integrators were derived using different splittings of the flow of the continuous Langevin dynamics. In our previous numerical tests\cite{Davidchack15jcp} we identified that Langevin A and C are more accurate than Langevin B in computing configurational quantities. \our{(Note also that, in the case of translational degrees of freedom, Langevin~C from Ref.~\onlinecite{Davidchack15jcp} coincides with the scheme called `BAOAB' in Ref.~\onlinecite{BenM13}, which was shown there to be the most accurate scheme among various types of splittings of Langevin equations for systems without rotational degrees of freedom and without HI.  The superior accuracy of this splitting is also demonstrated in the integration of Langevin dynamics with holonomic constraints~\cite{Leimkuhler16rspa}.)}
It was then natural to try to generalise Langevin A and C to the case considered here, the SDEs (\ref{la1})-(\ref{la2}). Using the same splitting as for Langevin C in Ref.~\onlinecite{Davidchack15jcp}, we have succeeded in constructing the presented above method (\ref{langC}) with the desirable properties, in particular that it is of second weak order. However, an attempt to generalize Langevin A failed, which is an interesting observation from the point of view of stochastic geometric integration.

We recall\cite{MT1} that weak-sense numerical methods for SDEs are sufficient for approximating expectations of the SDEs solutions, such as those considered in examples of Section~\ref{sec:ne2} \Rf{and Section~S2 in Supplementary Material}. When one aims to visualize individual trajectories of SDEs solutions (e.g., as in the example of Section~\ref{sec:ne3}), then mean-square (strong-sense) approximations are needed as they can ensure closeness of an approximate trajectory to the corresponding exact trajectory\cite{MT1}.  The proposed method (\ref{langC}) with random variables involved being simulated as $\mathcal{N}(0,1)$ is of mean-square order one, which is proved by comparing (\ref{langC}) with the mean-square Euler scheme and by applying the fundamental mean-square convergence theorem\cite{MT1}. We also note that often, as in the example of Section~\ref{sec:ne3}, the noise intensity is small. If we denote by $\varepsilon$ the parameter characterizing smallness of noise in (\ref{la1})-(\ref{la2}), then the mean-square accuracy of the method (\ref{langC}) with random variables simulated as $\mathcal{N}(0,1)$ is $O(h^2+\varepsilon h)$. The corresponding proof rests on the results from Ref.~\onlinecite{MT97sisc} (see also Ref.~\onlinecite[Chapter 3]{MT1}).

We point out, however, that even in the dynamical context we are not primarily concerned with the behaviour of individual trajectories \cite{FrenkelBook}. Nonlinearities lead to an exponential divergence of trajectories from those obtained in the  $h \rightarrow 0$ limit, and it is infeasible to compare individual trajectories directly with experiment due to a sensitive dependence on initial conditions \cite{FrenkelBook}. Instead, we are primarily interested in statistical properties of trajectories, for which expectations are more relevant.  Thus, the main practical interest is in weak convergence and weak-sense approximations.

{To approximate} the model with time-dependent non-conservative forces 
{(see the comment at the end of Section~\ref{sec:new})}, replace in (\ref{langC}):
\begin{itemize}
\item $\mathbf{f}(\mathbf{R}_{k},\mathbf{Q}_{k})$ by $\mathbf{f}(\mathbf{R}_{k},\mathbf{Q}_{k})+\tilde{\mathbf{f}}(t_{k},\mathbf{R}_{k},%
\mathbf{Q}_{k});$

\item $F^{j}(\mathbf{R}_{k},\mathbf{Q}_{k})$ by $F^{j}(\mathbf{R}_{k},\mathbf{Q}_{k})+\tilde{F}^{j}(t_{k},\mathbf{R}_{k},\mathbf{Q}_{k})$;

\item $\mathbf{f}(\mathbf{R}_{k+1},\mathbf{Q}_{k+1})$ by $\mathbf{f}(\mathbf{R}_{k+1},\mathbf{Q}_{k+1})+\tilde{\mathbf{f}}(t_{k+1},\mathbf{R}_{k+1},\mathbf{Q}_{k+1})$;

\item $F{^{j}}(\mathbf{R}_{k+1},\mathbf{Q}_{k+1})$ by $F{^{j}}(\mathbf{R}%
_{k+1},\mathbf{Q}_{k+1})+\tilde{F}(t_{k+1},\mathbf{R}_{k+1},\mathbf{Q}%
_{k+1}) $.
\end{itemize}
The resulting method is again of weak order $2$ and of mean-square order $1$.

\section{Numerical Experiments
\label{sec:num}}
\Rf{In order to test and explore the potential of the new integrator, we perform four sets of numerical experiments.  In the Supplementary Material, we describe two basic experiments that verify the correctness of the implementation of the integrator. First (see Section~S1), we demonstrate that the integrator without noise (i.e., $T = 0$) correctly converges to known analytical results for the dynamical properties of two sedimenting spheres. This study confirms the correct incorporation of hydrodynamic drag terms into the known equations for the dynamics of rigid bodies as represented by quaternions.\cite{Miller02}  Second (see Section~S2), we simulate Lennard-Jones spheres under periodic boundary conditions, where we test convergence and accuracy of the integrator with regard to its sampling from the canonical ensemble at a given temperature $T > 0$.
{The experiment in S2 is simply intended to show that the proposed Langevin-type equations and numerical integrator (incorporating potential forces, noise and hydrodynamic interactions) sample from the Gibbs distribution in steady state, as intended. We emphasize that we are not advocating for the method as an efficient sampler of the canonical ensemble -- the approach proposed in this paper is developed for simulating dynamics of hydrodynamically interacting rigid bodies.  If one needs just to sample from the canonical ensemble of rigid bodies, then the numerical integrators Langevin~A or Langevin~C from Ref.~\onlinecite{Davidchack15jcp} should be used. We note in passing that Langevin~C of Ref.~\onlinecite{Davidchack15jcp} is available in LAMMPS.}

Below, we describe two numerical experiments that demonstrate performance of the integrator on two model systems with HI that were previously investigated only via Brownian dynamics ({\it i.e.,} in the over-damped limit)\cite{ReichertDiss,Reichert04pre,Martin06prl}.  In Section~\ref{sec:ne2} we experiment with two spheres trapped in translational and rotational harmonic wells, and in Section~\ref{sec:ne3} we study circling spheres driven by external non-conservative force and torque. In the latter case, we are also able to use our integrator to extend the study to explore the coupling between rotational and translation effects, and the consequences of non-zero temperature.}

In all numerical experiments, the HYDROLIB package\cite{hydrolib} is used to calculate the hydrodynamic friction tensor $\xi(\mathbf{r})$.  It allows the calculation of HI in systems of equal radius spheres without boundaries (as in experiments described in \Rf{Section S1 of the Supplementary Material and Sections \ref{sec:ne2} and \ref{sec:ne3}) or with periodic boundary conditions (as in Section S2 of the Supplementary Material}).  The calculations are based on a multipole expansion \cite{Cichocki94} up to an order specified by integer parameter $l_\text{max}$, which can take values between 0 and 3.  We use $l_\text{max} = 2$ in our calculations.  Among other options available in HYDROLIB, we enable short-range corrections calculated from lubrication theory and use double-precision option for all external library functions.

\begin{figure}[tbp]
  \includegraphics[width=4cm]{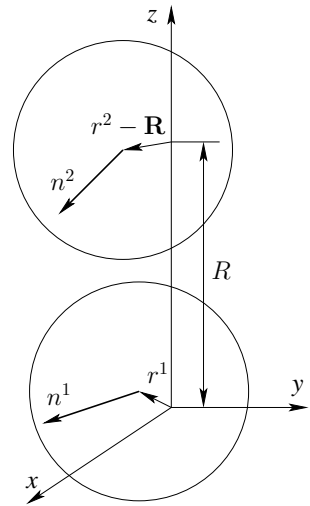}
\caption{\label{fig:twosph} Two harmonically trapped spheres.}
\end{figure}

\subsection{Two spheres trapped in translational and rotational harmonic wells\label{sec:ne2}}
In this section, we explore the time-correlation functions of two harmonically trapped spheres.  Their correlated motion arises exclusively due to hydrodynamic interactions. This setting allows us to demonstrate the dynamical consequences of both noise and HI as captured by the integrator, and the crossover to the over-damped limit. The set-up is similar to that in Refs.~\onlinecite{ReichertDiss,Reichert04pre,Martin06prl}.

The two spheres with coordinates $(r^i, q^i)$, $i = 1,2$, are illustrated in Fig.~\ref{fig:twosph}.  Unit vectors $n^i$ along the $x$ axis in the body-fixed coordinates of each sphere have space-fixed coordinates $n^i = (n^i_x, n^i_y, n^i_z)\tran = (A_{11}(q^i), A_{12}(q^i), A_{13}(q^i))\tran$, where $A_{kl}(q^i)$ are elements of the rotation matrix (\ref{eq:rotmat}).

Spheres 1 and 2 are trapped in translational harmonic wells at $(0, 0, 0)\tran$ and $\mathbf{R} = (0, 0, R)\tran$, respectively, as well as in rotational harmonic wells with respect to $n^i_x$. The potential energy of the system is thus given by (cf.~Eq.~(4.3) in Ref.~\onlinecite{ReichertDiss}):
\begin{equation} \label{eq:poten2}
  U({\bf r}, {\bf q}) = \frac{k^t}{2}(|r^1|^2 + |r^2 - \mathbf{R}|^2) - \frac{k^r}{2}\left[(n^1_x)^2 + (n^2_x)^2\right].
\end{equation}

The translational forces on the two spheres are
\begin{eqnarray} \label{eq:tforce}
  f^1 &=& -\nabla_{r^1} U = -k^t r^1\,,\nonumber\\
  f^2 &=& -\nabla_{r^2} U = -k^t(r^2 - \mathbf{R})\,.
\end{eqnarray}
The rotational forces, calculated according to
\begin{equation} \label{eq:qforce1}
 F^i = -\nabla_{q^i} U - (q^{i\mathsf T} \nabla_{q^i} U)q^i,
\end{equation}
$i = 1, 2$, take the form
\begin{equation} \label{eq:qforce2}
  F^i = 4k^r(q_0^{i2} + q_1^{i2} - q_2^{i2} - q_3^{i2})
  \left(\begin{array}{c}
  q_0^i(q_2^{i2} + q_3^{i2})\\
  q_1^i(q_2^{i2} + q_3^{i2})\\
 -q_2^i(q_0^{i2} + q_1^{i2})\\
 -q_3^i(q_0^{i2} + q_1^{i2})
  \end{array}\right).
\end{equation}

Guided by Ref.~\onlinecite{Reichert04pre}, we compute time-correlation functions (TCFs) among the following variables:
\begin{equation}\label{eq:tran}
\mbox{transversal modes:}\quad x^1, x^2, \chi_y^1, \chi_y^2,
\end{equation}
\begin{equation}\label{eq:long}
\mbox{longitudinal modes:}\quad z^1, \bar{z}^2, \chi_z^1, \chi_z^2,
\end{equation}
where $\bar{z}^2 = z^2 - R$
and $\chi_\alpha^i$ is the angle of rotation of $n^i$ about the $\alpha$-axis, so that $\tan \chi_y^i = -n_z^i/n_x^i$ and $\tan \chi_z^i = n_y^i/n_x^i$ (cf.~Eqs.~(6.3) and
(6.4) in Ref.~\onlinecite{ReichertDiss}).

We use the notation
\begin{equation}\label{eq:tcf}
  \langle \alpha, \beta \rangle = \frac{\langle \alpha(t+\tau)\beta(t)\rangle}
  {\sqrt{\langle\alpha^2(t)\rangle\langle\beta^2(t)\rangle}}
\end{equation}
for the TCF of $\alpha(t)$ and $\beta(t)$.

We computed the following TCFs for pairs of variables formed from the set in (\ref{eq:tran},\ref{eq:long})
(taking account of the system symmetry with respect to sphere numbering):
\begin{itemize}
\item Auto-correlations: $\langle x^1, x^1\rangle$, $\langle z^1, z^1\rangle$, $\langle \chi_y^1, \chi_y^1\rangle$, $\langle \chi_z^1, \chi_z^1\rangle$;
\item Cross-correlations (betweens two spheres): $\langle x^1, x^2\rangle$, $\langle z^1, \bar{z}^2\rangle$, $\langle \chi_y^1, \chi_y^2\rangle$, $\langle \chi_z^1,
    \chi_z^2\rangle$, $\langle x^1, \chi_y^2\rangle$;
\item Mixed self-correlations: $\langle x^1, \chi_y^1\rangle$.
\end{itemize}
All other pairs of variables formed from the set in (\ref{eq:tran},\ref{eq:long}) are uncorrelated (see the corresponding discussion in
Refs.~\onlinecite{Reichert04pre,ReichertDiss}).
The cross-correlations and mixed self-correlations are the results of the hydrodynamic interaction between the two spheres.  Note that mixed self-correlations are zero in the absence of the second sphere \cite{Reichert04pre}.

The following reduced units are imposed by the HYDROLIB package: the radius of the spheres is set to 1, the viscosity of the surrounding fluid $\eta$ is set to $1/(4\pi)$. \our{In addition, the mass scale is set by choosing the mass of the spheres $m=1.0$.}
Numerical experiments were carried out on the systems with the following parameters: $I = (0.4, 0.4, 0.4)$, $R = 3.0$ and $3.5$, $k^t = 10.0$, $k^r = 20.0$, temperatures $k_B T = 0.1$, $0.2$, and $0.5$.   We used a relatively small time step of $h = 0.002$ in all simulations.

\begin{figure}[tbp]
  \includegraphics[width=8.5cm]{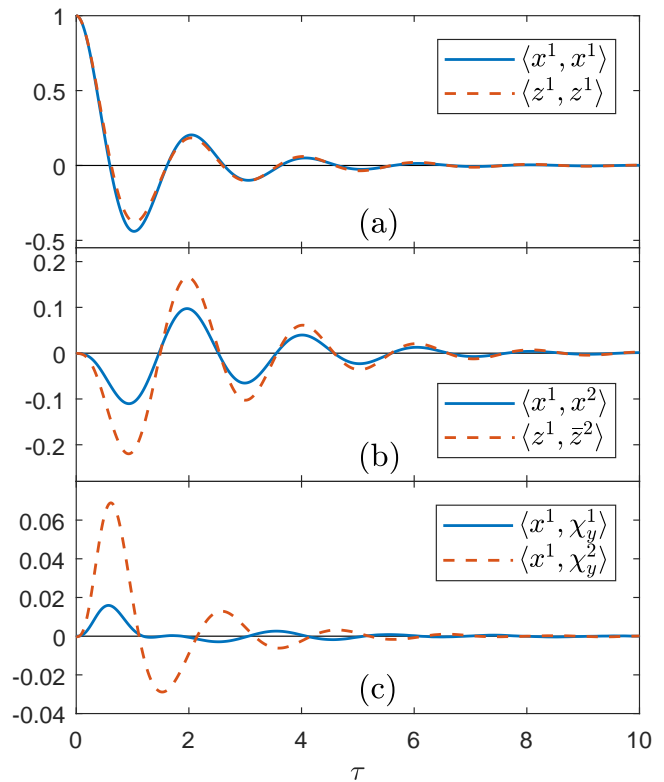}
  \caption{\label{fig:tcf_xz} Translational-translational and translational-rotational TCFs.  System parameters: $m = 1.0$, $I = (0.4, 0.4, 0.4)$, $R = 3.0$, $k^t = 10.0$, $k^r = 20.0$, $k_B T = 0.1$.
  (a) Translational auto-correlation functions. The difference between the ACFs for longitudinal and transversal modes is due to hydrodynamic interactions;
  (b) Translational cross-correlation functions;
  (c) Mixed cross-correlation and self-correlation functions.  As expected, the self-correlation effect (solid blue line) is weaker than cross-correlation because it is induced by the presence of the second sphere.}
\end{figure}

\begin{figure}[tbp]
  \includegraphics[width=8.5cm]{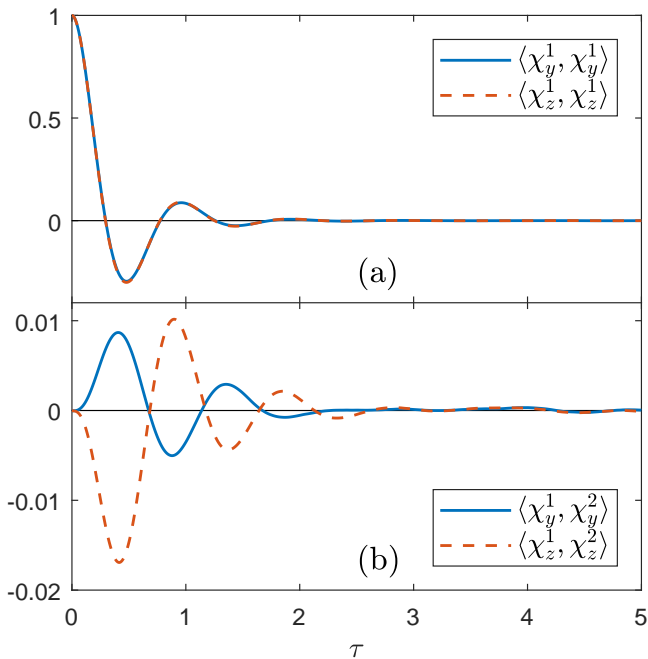}
  \caption{\label{fig:tcf_ny} Rotational TCFs with the same system parameters as in Figure~\ref{fig:tcf_xz}.  (a) Rotational ACFs. The difference between the ACFs of the rotational longitudinal and transversal modes is very small. (b) Rotational cross-correlation functions.}
\end{figure}

Figures~\ref{fig:tcf_xz}(a) and \ref{fig:tcf_ny}(a) show translational and rotational auto-correlation functions (ACFs).  Compared to over-damped dynamics, in which auto-correlations exhibit monotonic exponential decay~\cite{Reichert04pre}, auto-correlations in the Langevin dynamics exhibit decaying oscillations which are characteristic of the inertial effects. Crucially, however, we observe the expected cross-correlations and mixed self-correlations with time delay~\cite{Reichert04pre}. Measurements of the translational and rotational kinetic energies of the spheres confirm thermal equilibration of the system at the correct temperature.  We also observe that $\langle (x^i)^2\rangle = \langle (z^i)^2 \rangle = (k_B T)/k^t$, as expected from the equipartition theorem.  The shape of the TCFs shows very weak temperature dependence, while the magnitude of cross-correlations and mixed self-correlations decreases with increasing $R$.

\begin{figure}[tbp]
  \includegraphics[width=8.5cm]{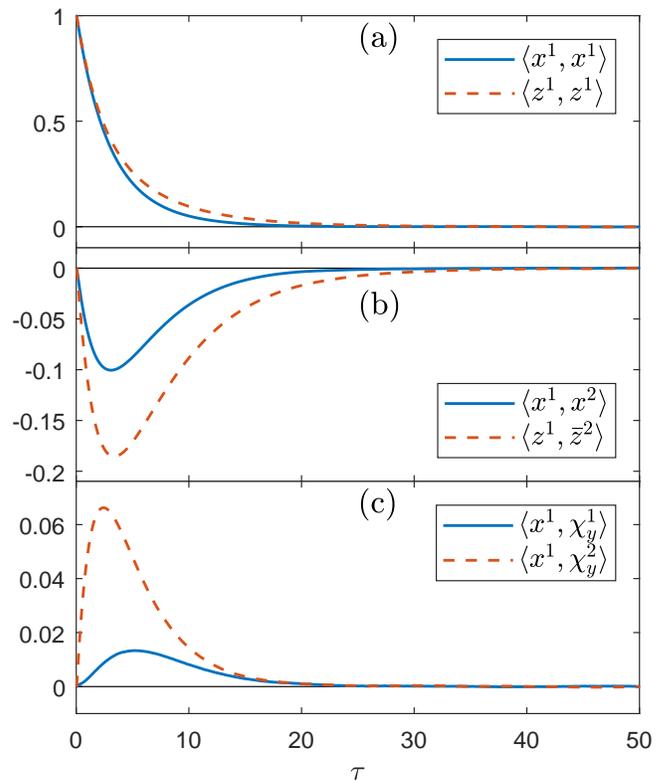}
  \caption{\label{fig:tcf_xz_highv} Same as Figure~\ref{fig:tcf_xz} for the two-sphere system with parameters $m = 0.05$, $I = (0.02, 0.02, 0.02)$, $R = 3.0$, $k^t = 0.5$, $k^r =
  1.0$, $k_B T = 0.005$.}
\end{figure}

\begin{figure}[tbp]
  \includegraphics[width=8.5cm]{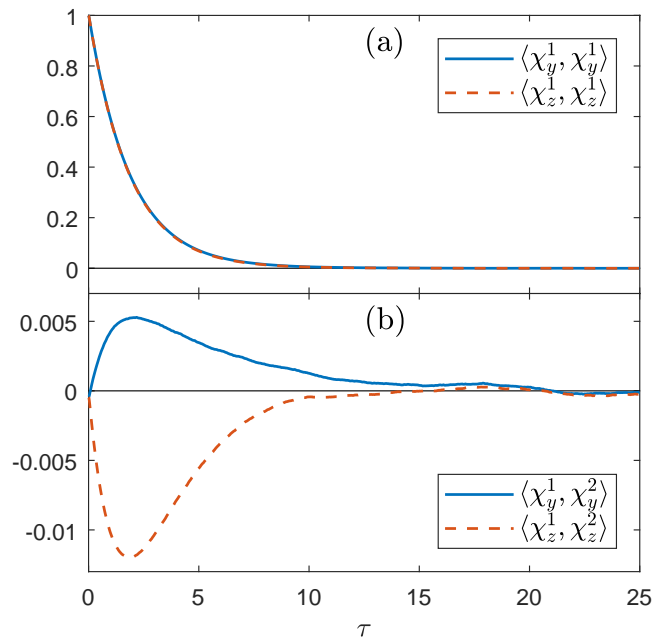}
  \caption{\label{fig:tcf_ny_highv} Same as Figure~\ref{fig:tcf_ny} for the two-sphere system with parameters $m = 0.05$, $I = (0.02, 0.02, 0.02)$, $R = 3.0$, $k^t = 0.5$, $k^r =
  1.0$, $k_B T = 0.005$.}
\end{figure}

Because the proposed numerical integrator (\ref{langC})  exactly solves the Ornstein-Uhlenbeck part (\ref{ou4}) of the Langevin equations (\ref{la1})-(\ref{la2}), it can be applied in any hydrodynamic viscosity regime, including high viscosity, where an over-damped dynamical model is typically used.  To illustrate this, we have applied our integrator to a system with smaller mass/inertia and energy/temperature, which corresponds to higher viscosity.  In Figures~\ref{fig:tcf_xz_highv} and \ref{fig:tcf_ny_highv} we show the results for the system with parameters which correspond to 20 times higher viscosity than that shown in Figures~\ref{fig:tcf_xz} and \ref{fig:tcf_ny}.  The integration time step is $h = 0.01$.  Excellent qualitative agreement of the TCFs with those modelled in Ref.~\onlinecite{Reichert04pre} using over-damped dynamics or in experimental measurements~\cite{Meiners99prl,Bartlett01rsta,Martin06prl} is observed. In particular, we note the presence of clearly ``anti-correlated'' behaviour (see negative TCFs in Figure~\ref{fig:tcf_xz_highv}(b)), arising from a resistance to shearing or changing the volume of fluid between the two spheres \cite{Reichert04pre}.

\subsection{Circling spheres driven by external force and torque\label{sec:ne3}}
\begin{figure}[tbp]
  \includegraphics[width=5cm]{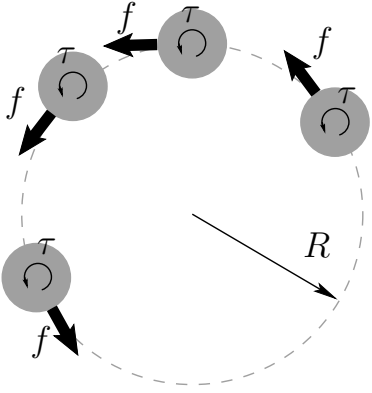}
\caption{\label{fig:ring} Spheres circling around a ring of radius $R$ pushed by tangential force $f$ and torques $\tau$.}
\end{figure}

Here we demonstrate application of our integrator to a system driven by non-conservative forces. The setup, shown in Figure~\ref{fig:ring}, is similar to that in Ref.~\onlinecite{ReichertDiss}. Spheres with coordinates $r^i = (x^i, y^i, z^i)\tran$ are placed around a ring of radius $R$ in the $x$-$y$ plane tethered by a radial harmonic potential and a harmonic potential along the $z$ axis
\begin{equation}\label{eq:rad}
    U_\mathrm{rad} = \frac{k^t}{2} \sum_{i=1}^N [(\rho^i-R)^2 + (z^i)^2],
\end{equation}
where $N$ is the number of spheres and $\rho^i = \sqrt{(x^i)^2 + (y^i)^2}$. The spheres interact with one another via a short-range repulsive potential preventing their overlap
\begin{equation}\label{eq:rep}
   U_\mathrm{rep} = \sum_{i=2}^N\sum_{j=1}^{i-1} A\left[\left(\frac{|r^i-r^j|}{2}\right)^{12} - 1\right]^{-1}.
\end{equation}
The repulsive potential is smoothly truncated at the cut-off distance of 2.4 reduced units.  In addition, a force with magnitude $f$ is applied to each sphere in the direction tangent to the ring: $(-f y^i/\rho^i, f x^i/\rho^i, 0)\tran$ and torque with magnitude $\tau$ is applied to spin each sphere in the $x$-$y$ plane: $(0, 0, \tau)\tran$.
\begin{figure}[tbp]
\begin{center}
  \includegraphics[width=8.5cm]{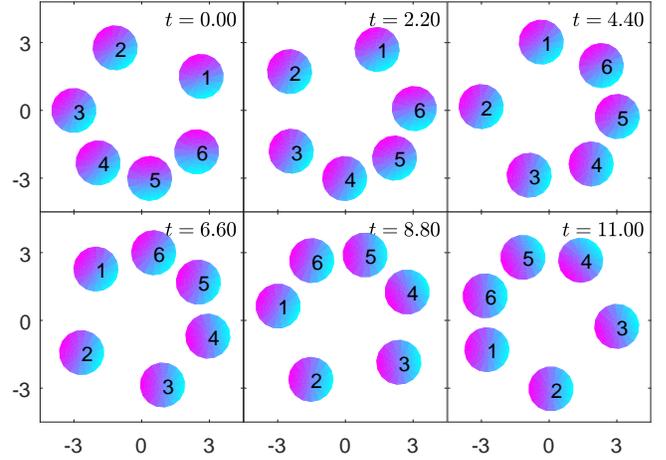}
\end{center}
\caption{\label{fig:rs1} Illustration of the limit cycle exhibiting drafting effect. $f = 1.0$, $\tau = 0$, $T = 0$.  Shading on the spheres indicates their orientation.  Multimedia view: 
\href{https://www.dropbox.com/s/cgbs1g7dri2w5zk/ring_f1.0tau0T0.mp4?dl=0}{\tt ring\_f1.0tau0T0.mp4}}
\end{figure}

In Ref.~\onlinecite{ReichertDiss}, only noiseless (i.e., $T=0$) translational dynamics was investigated.  Our numerical integrator allows us to investigate the coupling between translational and rotational motion of the spheres, and non-zero temperature effects.  Here we present an example of a system of $N = 6$ spheres on a ring of radius $R = 3.0$.  The mass and moments of inertia of each sphere are $m = 1.0$, $I = (0.4, 0.4, 0.4)$. The potential energy constants in Eqs.~(\ref{eq:rad}) and (\ref{eq:rep}) are $k^t = 10.0$ and $A = 0.02$, corresponding to relatively strong radial trap and short-range repulsion.  We investigate the dependence of the dynamics of this system on parameters $f$, $\tau$, and $T$. Since we are interested in visualizing spheres' trajectories, we rely on the mean-square convergence of the method (\ref{langC}) appropriately augmented with non-conservative forces (see the end of Section~\ref{sec:met}).

For $f > 0$, $\tau = 0$, and $T = 0$, the spheres move around the ring anticlockwise in a limit cycle exhibiting a drafting effect\cite{ReichertDiss}, where a cluster of five spheres moves faster and catches up the sixth sphere, while the trailing sphere in the cluster gets dropped, as shown in Figure~\ref{fig:rs1}.  Due to the hydrodynamic coupling between translational and rotational degrees of freedom, the spheres also spin anticlockwise in the $x$-$y$ plane (as seen by the shading of the spheres).  The sphere velocities around the ring and the spin angular velocities increase with increasing $f$. Note that the induced rotation of the spheres in turn influences the translational velocity around the ring, emphasising the importance of considering rotational motion even in the absence of the torque ($\tau=0$).

\begin{figure}[tbp]
\begin{center}
  \includegraphics[width=8.5cm]{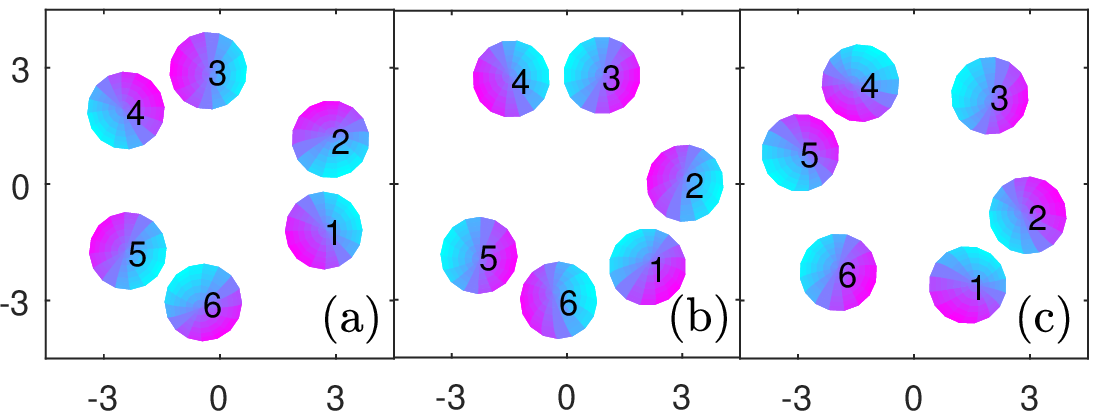}\\
\end{center}
\caption{\label{fig:rs2} Illustration of other limiting behaviours of the system with $\tau < 0$.  (a) Relative steady state: $f = 1.2$, $\tau = -4.0$.  Multimedia view: \href{https://www.dropbox.com/s/amnenj18shzy02f/ring_f1.2tau-4.0T0.mp4?dl=0}{\tt ring\_f1.2tau-4.0T0.mp4}; (b) pair drafting limit cycle: $f = 1.2$, $\tau = -3.0$.  Multimedia view: \href{https://www.dropbox.com/s/xrziqd7fs0ms892/ring_f1.2tau-3.0T0.mp4?dl=0}{\tt ring\_f1.2tau-3.0T0.mp4}; (c) limit cycle with two triplets drafting: $f = 0.4$, $\tau = -4.0$. Multimedia view: \href{https://www.dropbox.com/s/331d57dl1hwmcn9/ring_f0.4tau-4.0T0.mp4?dl=0}{\tt ring\_f0.4tau-4.0T0.mp4}.}
\end{figure}

The drafting limit cycle persists for $\tau > 0$, with the spheres spinning faster and moving faster anticlockwise around the ring.  However, when $\tau < 0$, the spheres are pushed to spin clockwise, causing disruption of the limit cycle through hydrodynamic interactions and the emergence of other asymptotic behaviours, shown in Figure~\ref{fig:rs2}. The most common is the relative steady state solution (i.e. relative with respect to the orbital rotation symmetry), shown in Figure~\ref{fig:rs2}(a), where spheres move in pairs around the ring at fixed distances, constant spin angular velocities (clockwise), and constant orbiting velocity.

Other types of limit cycles are observed in the $(f, \tau)$ parameter plane, usually on the boundaries between the drafting limit cycle in Figure~\ref{fig:rs1} and the steady state solution in Figure~\ref{fig:rs2}(a).  For $f = 0$ and $\tau \neq 0$, a different type of drafting limit cycle is observed, where the spinning of the spheres induces, through the hydrodynamic coupling, orbital motion around the ring \our{(see Supplementary Material: file \href{https://www.dropbox.com/s/b6g8rpo0ggldgq6/ring_f0tau-5.0aT0.mp4?dl=0}{\tt ring\_f0tau-5.0aT0.mp4})}.  At relatively large $f$ and somewhat weaker negative $\tau$ compared to the steady state solution, we observe a drafting limit cycle shown in Figure~\ref{fig:rs2}(b), where a pair of trailing spheres detaches form the back of the faster moving cluster of four spheres and then is recaptured while another pair of spheres is dropped at the back.  For weaker $f$ and relatively strong $\tau$, yet another type of a drafting limit cycle is observed where spheres move in triplets as shown Figure~\ref{fig:rs2}(c): spheres are orbiting clockwise with spheres 3 and 6 getting caught by faster moving pairs 4,5 and 2,1, respectively, followed by spheres 5 and 2 getting dropped at the back of the moving triplets.

With $T > 0$, noise is introduced into the system. Different types of limit cycles exhibit different degrees of sensitivity to noise.  For $k_B T \lesssim 0.0001$, all limit cycles discussed above can still be observed, while for $k_B T \gtrsim 0.001$ we mostly observe the limit cycle shown in Figure~\ref{fig:rs1} and the steady state shown in Figure~\ref{fig:rs2}(a) \our{(Supplementary Material: files \href{https://www.dropbox.com/s/v1e7mlalkaivrnt/ring_f1.0tau0T0.01.mp4?dl=0}{\tt ring\_f1.0tau0T0.01.mp4} and \href{https://www.dropbox.com/s/n0r2po44qh6dw5y/ring_f1.2tau-4.0T0.01.mp4?dl=0}{\tt ring\_f1.2tau-4.0T0.01.mp4})}.  Close to the boundary between the two solutions in the $(f, \tau)$ parameter plane, we also observe the system switching between the two solutions at random time intervals.  Such behaviour can be characterised as noise-induced intermittency\cite{Eckmann81jpa}.

\section{Conclusions}\label{sec:conc}
We have proposed and tested a quaternion-based geometric integrator for Langevin SDEs that incorporates cooperative hydrodynamic interactions between rigid bodies. The integrator takes a user-defined approximation to the multi-particle friction tensor as an input, does not assume the over-damped limit, and is of second order in the weak sense. Further, the integrator naturally conserves quaternion length and is symplectic in the noise-free, friction-free limit. To our knowledge, this is the first quaternion-based integrator incorporating cooperative hydrodynamics that has been implemented and tested.

Langevin dynamics with hydrodynamic interactions (so-called Stokesian dynamics) is widely used to understand a range of systems, including the rheology of colloidal suspensions and cellular transport \cite{Dlugosz2011,Skolnick2016,Park2016}. Our method will facilitate the incorporation of rotational motion into these descriptions, which is important for modelling, e.g. patchy colloids \cite{Glotzer2007,Pawar2010,Chen2011,Newton2015} and globular proteins \cite{Dlugosz2011,Skolnick2016}.


Historically, Stokesian integrators have assumed an over-damped (``Brownian'') limit in which the inertia of the  simulated particles is neglected \cite{SnookBook,Ermak1978,Dickinson1985,Brady1988,Ilie2015}. By contrast, our algorithm explicitly retains the generalised momenta; the over-damped limit can be approached simply by setting friction coefficients to large values. This setup allows the construction of a Verlet-like integrator with second-order weak accuracy in the time-step. We expect that this approach will be  particularly useful in two contexts. Firstly, it will allow the study of systems that are not over-damped; \our{as illustrated in Section~\ref{sec:ne2},} such systems can show substantially different behaviour from their over-damped counterparts \cite{Ai2017}.

Alternatively, in systems with stiff interactions, it is important to accurately integrate the potential forces that contribute to the equations of motion by using small time steps. As was observed, for example, in Ref.~\onlinecite{Davidchack09,Davidchack15jcp}, the use of Brownian dynamics is often hindered by inefficiency of numerical integrators which, under the requirement of a single evaluation of the forces and friction matrix per step of an algorithm, are only of weak first-order accuracy in comparison with second-order weak numerical schemes available for Langevin systems.  Consequently, the combination of Langevin equations and a second-order geometric integrator is usually more computationally efficient than the combination of Brownian dynamics and a first-order scheme. We therefore expect that the numerical method proposed in this paper will be a powerful tool for studying the over-damped limit, just as similar second-order integrators (without HI) have been successfully used to simulate coarse-grained models with stiff potential functions \cite{Takagi2003,Ouldridge_walker_2013,Machinek2014,Hinckley2014}.

{We find that the dominant contribution to the computational cost of our simulations is the HYDROLIB-based calculation of $\xi({\bf r},{\bf q})$. Such a calculation, or a similar one to obtain $\mu({\bf r},{\bf q}) = \xi({\bf r},{\bf q})^{-1}$, is not part of our method {\it per se}, but is fundamental to Stokesian dynamics. Similarly, the costs of calculating forces and obtaining a noise covariance matrix by decomposing $\xi({\bf r},{\bf q})$ or  $\mu({\bf r},{\bf q})$ are inherent in any Stokesian method. The operations unique to our method -- the actual details of the coordinate updates in (\ref{langC}) -- are essentially irrelevant to the computational cost. 
{In terms of computational efficiency, the key feature of our integrator is that it requires only one evaluation of the friction matrix and other forces per timestep.}

The computationally intensive Stokesian approach is particularly suited to studying small systems using accurate hydrodynamic models; depending on the accuracy required, larger systems may be better treated by methods such as Lattice Boltzmann\cite{Ladd1993}, Dissipative Particle Dynamics\cite{Groot1997} or Multiparticle Collision Dynamics\cite{Malevanets1999}, which use a simplified but explicit representation of the solvent.
In-depth discussions of the relative merits of different approximations can be found elsewhere\cite{Skolnick2016}. However, our specific method opens up clear possibilities in several contexts.} In particular, we note that models of externally-driven  colloids or self-propelled swimmers rely on the hydrodynamic interaction between only a few bodies to produce interesting phenomena \cite{Najafi2004,Dunkel2009,ReichertDiss}. Some work has been done to understand the potential importance of noise for swimmers \cite{Dunkel2009,Lauga2011}; our integrator allows the incorporation of both rotational motion and inertia into this perspective.  As we have shown in the case of spheres driven on a circular path, 
{described in Section~\ref{sec:ne3}, the interplay between hydrodynamic drafting and rotational motion leads to the formation of previously unobserved
dynamical patterns (e.g., steady orbital motion of pairs of spinning spheres).}

{Another obvious use of our integrator} is in studying self-assembly.  Recently, several authors have highlighted the important role of diffusion in determining the self-assembly pathways of finite-sized structures in the absence of cooperative hydrodynamics \cite{Newton2015,Vijaykumar2016}. Our integrator would enable this analysis to be extended to self-assembling systems with hydrodynamic interactions.

\section*{Supplementary Material}
\Rf{Supplementary Material describes additional numerical experiments in Sections S1 and S2. Section S3 contains the list of videos of the dynamical behaviour of spheres moving on a ring under the influence of orbital force and a spinning torque, as discussed in Section~\ref{sec:ne3}. Implementation details of the numerical integrator are given in Section S4.}

\section*{Acknowledgment}
This work was partially supported by the Computer Simulation of Condensed Phases (CCP5) Collaboration Grant, which is part of the EPSRC grant EP/J010480/1. T.E.O. is supported by a Royal Society University Research Fellowship, and also acknowledges fellowships from University College, Oxford and Imperial College London.  R.L.D. acknowledges a study leave granted by the University of Leicester. This research used the ALICE High Performance Computing Facility at the University of Leicester.

\appendix{}
\section{Evaluation of covariance integral and matrix $\sigma$ \label{app:a}}
Let us rewrite $\tilde{\xi}(\mathbf{r},\mathbf{q})$ \our{from (\ref{ou4})} as
\begin{equation} \label{tk1}
\tilde{\xi}(\mathbf{r},\mathbf{q}) = G_{1}(\mathbf{q}) \xi(\mathbf{r},\mathbf{q}) G_{2}(\mathbf{q}),
\end{equation}
where the $7n$-by-$6n$ matrix $G_{1}$ and the $6n$-by-$7n$ matrix $G_{2}$ have the following block structure
\begin{equation}\label{tk2}
G_{1}(\mathbf{q})=\left[\begin{array}{cc}
\bm{1}_{3n} & \bm{0} \\
\bm{0} & G_{11}(\mathbf{q})%
\end{array}\right], \ \
G_{2}(\mathbf{q})=\left[\begin{array}{cc}
G_{21} & \mathbf{0} \\
\mathbf{0} & G_{22}(\mathbf{q}) %
\end{array}%
\right].
\end{equation}%
Here $\mathbf{0}$ are zero matrices of the corresponding dimensions, $\bm{1}_{3n}$ is the $3n$-dimensional identity matrix, $G_{11}$ is the $4n$-by-$3n$ block-diagonal matrix with diagonal $4$-by-$3$ blocks
\begin{equation}\label{tk3}
G_{11}^{(i,i)}(\mathbf{q}) = 2\check{S}(q^{i}),\ \ i=1,\ldots ,n,
\end{equation}
$G_{21}$ is the $3n$-by-$3n$ diagonal matrix with the $3$-by-$3$ diagonal blocks
\begin{equation} \label{tk4}
G_{21}^{(j,j)}=\frac{1}{m^{j}}\bm{1}_3,\ \ j=1,\ldots ,n,
\end{equation}
and $G_{22}$ is the $3n$-by-$4n$ block-diagonal matrix with the diagonal $3$-by-$4$ blocks
\begin{equation} \label{tk5}
G_{22}^{(j,j)}=\frac{1}{2} A^{\mathsf{T}}(q^{j})\hat{D}^{j}\hat{S}^{\mathsf{T}}(q^{j}),\ \
j=1,\ldots ,n.
\end{equation}%
We can also rewrite $\tilde{b}(\mathbf{r},\mathbf{q})$ \our{from (\ref{ou4})} as
\begin{equation}  \label{tk6}
\tilde{b}(\mathbf{r},\mathbf{q}) = G_{1}(\mathbf{q})b(\mathbf{r},\mathbf{q})
\end{equation}%
and thus, using Eq.~(\ref{eq:bbxi}),
\begin{equation} \label{tk7}
\tilde{b}(\mathbf{r},\mathbf{q})\tilde{b}^{\mathsf{T}}(\mathbf{r},\mathbf{q})
= \frac{2}{\beta}G_{1}(\mathbf{q})\xi(\mathbf{r},\mathbf{q})G_{1}^{\mathsf{T}}(\mathbf{q}).
\end{equation}%
In addition we define the $6n$-by-$6n$ matrix
\begin{equation} \label{tk8}
K:=G_{2}G_{1}=\left[\begin{array}{cc}
G_{21} & \bm{0} \\
\bm{0} & K_2(\mathbf{q})
\end{array}\right],
\end{equation}
where $K_2 := G_{22}(\mathbf{q})G_{11}(\mathbf{q})$.  The $3$-by-$3$ blocks on the main diagonal of $K_2$ have the form
\begin{equation} \label{tk10}
A^{\mathsf{T}}(q^{j})\hat{D}^{j}\hat{S}^{\mathsf{T}}(q^{j})\check{S}%
(q^{j})=A^{\mathsf{T}}(q^{j})\hat{D}^{j}A(q^{j}).
\end{equation}
Therefore, both $K_2$ and $K$ are symmetric:
\begin{equation} \label{tk12}
K=K^{\mathsf{T}}=G_{1}^{\mathsf{T}}G_{2}^{\mathsf{T}}.
\end{equation}

Using the definition of the matrix exponent and properties of the above defined matrices, we \our{obtain}
\begin{equation*}
\mathrm{e}^{-\tilde{\xi}(t-s)}G_{1}\xi = G_{1} \xi \mathrm{e}^{-K\xi (t-s)}
\end{equation*}
as well as
\begin{equation*}
G_{1}^{\mathsf{T}}\mathrm{e}^{-\tilde{\xi}^{\mathsf{T}}(t-s)} =
\mathrm{e}^{-K\xi (t-s)}G_{1}^{\mathsf{T}}.
\end{equation*}
Thus, the covariance integral in Eq.~(\ref{ou6}) can be evaluated as follows
\begin{eqnarray*}
C(t;\mathbf{r},\mathbf{q}) &=& \int_{0}^{t}\mathrm{e}^{-\tilde{\xi}(\mathbf{r},\mathbf{q})(t-s)}\tilde{b}(\mathbf{r},\mathbf{q})\tilde{b}(\mathbf{r},\mathbf{q})^{\mathsf{T}}\mathrm{e}^{-\tilde{\xi}(\mathbf{r},\mathbf{q})^{\mathsf{T}}(t-s)}ds \\
&=&\frac{2}{\beta }G_{1}(\mathbf{q})\xi(\mathbf{r},\mathbf{q}) \left[ \int_{0}^{t}\mathrm{e}^{-2K(\mathbf{q})\xi(\mathbf{r},\mathbf{q})(t-s)}ds\right] G_{1}^{\mathsf{T}}(\mathbf{q}) \\
&=&\frac{1}{\beta }G_{1}(\mathbf{q})\xi(\mathbf{r},\mathbf{q}) \left[ K(\mathbf{q})\xi(\mathbf{r},\mathbf{q}) \right]^{-1}\\
&&\times \left[\bm{1}_{6n}-\mathrm{e}^{-2K(\mathbf{q}) \xi(\mathbf{r},\mathbf{q}) t}\right] G_{1}^{\mathsf{T}}(\mathbf{q}) \\
&=&\frac{1}{\beta }G_{1}(\mathbf{q})K^{-1}(\mathbf{q})\left[\bm{1}_{6n} - \mathrm{e}^{-2K(\mathbf{q})\xi (\mathbf{r},\mathbf{q})t}\right] G_{1}^{\mathsf{T}}(\mathbf{q}).
\end{eqnarray*}%
Note that
\begin{equation}\label{tk18}
K^{-1}(\mathbf{q})=\left[\begin{array}{cc}
G_{21}^{-1} & \mathbf{0} \\
\mathbf{0} & K_2^{-1}(\mathbf{q})%
\end{array}\right],
\end{equation}%
which is easy to compute analytically since $G_{21}$ is diagonal and $K_2(\mathbf{q})$ is block diagonal with blocks on the diagonal being $A^{\mathsf{T}}(q^{j})\hat{D}^{j}A(q^{j})$ and
\begin{equation}\label{tk19}
\left[ A^{\mathsf{T}}(q^{j})\hat{D}^{j}A(q^{j})\right] ^{-1}=A^{\mathsf{T}}(q^{j})\left[ \hat{D}^{j}\right] ^{-1}A(q^{j}).
\end{equation}

We can now write the matrix $\sigma(t;\mathbf{r},\mathbf{q})$ in Eq.~(\ref{lbec}) in the form
\begin{equation} \label{tk21}
\sigma (t;\mathbf{r},\mathbf{q})=\frac{1}{\sqrt{\beta }}G_{1}(\mathbf{q})\tilde{\sigma}(t;\mathbf{r},\mathbf{q}),
\end{equation}
where the $6n$-by-$6n$ matrix $\tilde{\sigma}(t;\mathbf{r},\mathbf{q})$ satisfies
\begin{equation} \label{tk22}
\tilde{\sigma}(t;\mathbf{r},\mathbf{q})\tilde{\sigma}^{\mathsf{T}}(t;\mathbf{r},\mathbf{q}) = K^{-1}(\mathbf{q})\left[\bm{1}_{6n} - \mathrm{e}^{-2K(\mathbf{q})\xi (\mathbf{r},\mathbf{q})t}\right]
\end{equation}
which can be computed by Cholesky factorization.

We also note that the matrix exponent $\mathrm{e}^{-\tilde{\xi}(\mathbf{r},\mathbf{q})t}$ from (\ref{ou7}), which is used in the method (\ref{langC}), can be expressed as
\begin{eqnarray*}
\mathrm{e}^{-\tilde{\xi}(\mathbf{r},\mathbf{q})t}
&=&G_{1}(\mathbf{q})\left[ \mathrm{e}^{-K(\mathbf{q})\xi (\mathbf{r},\mathbf{q}%
)t}\right] ^{\mathsf{T}}\ G_{1}^{\mathsf{T}}(\mathbf{q})\left[
\begin{array}{cc}
\mathbf{\bm{1}}_{3n} & \mathbf{0} \\
\mathbf{0} & \bm{1}_{4n}/4%
\end{array}%
\right] \\
&&+\left[
\begin{array}{cc}
\mathbf{0} & \mathbf{0} \\
\mathbf{0} & \mathbf{qq}^{\mathsf{T}}%
\end{array}%
\right],
\end{eqnarray*}
where $\mathbf{qq}^{\mathsf{T}}$ means the block diagonal matrix with $4\times 4$ blocks being $(q^{j})(q^{j})^{\mathsf{T}}$. That is, per step of (\ref{langC}) we only need to compute one matrix exponent, $\mathrm{e}^{-K(\mathbf{q})\xi (\mathbf{r},\mathbf{q})t}$. However, in
{Section S4 of the Supplementary Material}
we present, for better clarity, the implementation with two matrix exponents.

%

\clearpage

\begin{widetext}
\begin{center}
\textbf{\large 
Supplementary Material for ``Geometric Integrator for Langevin Systems with Quaternion-based Rotational Degrees of Freedom and Hydrodynamic Interactions''}
\end{center}




{\bf Table of Contents}
\begin{itemize}
\item {\bf S1:} Asymptotic motion of two sedimenting spheres
\item {\bf S2:} Lennard-Jones spheres in periodic boundary conditions
\item {\bf S3:} Videos of circling spheres driven by external force and torque
\item {\bf S4:} Implementation details of the numerical integrator
\end{itemize}

\end{widetext}

\beginsupplement
\makeatletter
\renewcommand{\bibnumfmt}[1]{[S#1]}
\renewcommand{\citenumfont}[1]{S#1}

\noindent
{\bf Section S1: Asymptotic motion of two sedimenting spheres}\\[1ex]
In order to verify our use of HYDROLIB\cite{hydrolib_s} for computing HI and integration of the corresponding equations of motion in the noiseless regime (i.e., temperature $T = 0$), we compare our simulations to analytic results for the dynamics of two sedimenting spheres.

Analytic results for the dynamics of hydrodynamically-coupled rigid bodies are rare due to the inherent complexity of such systems. However, they exist for two spheres of radius $a$ separated by distance $l$ along the $x$ axis, undergoing sedimentation parallel to the $z$-axis (pointing down) due to gravity and without noise\cite{HappelBook}. Hydrodynamic coupling increases the sedimentation velocity and leads to rotation of the spheres.

Taking the sphere coordinates as $(-l/2, 0, 0)\tran$ and $(l/2, 0, 0)\tran$, and the gravitational force as $(0, 0, F)\tran$ on both spheres, Happel and Brenner\cite{HappelBook} provide expansions in powers of $a/l$ for the asymptotic velocity $(0, 0, U)\tran$ of both spheres, and angular velocities $(0, -\omega, 0)\tran$ and $(0, \omega, 0)\tran$, respectively (see Eqs.~(6-3.97) and (6-3.100) in Ref.~\onlinecite{HappelBook}):
\begin{eqnarray}\label{eq:sd1}
F = 6\pi \eta a U \left(1 - \frac{3}{4}\frac{a}{l} + \frac{9}{16}\frac{a^2}{l^2} - \frac{59}{64}\frac{a^3}{l^3}\right.\nonumber\\ \left. + \frac{273}{256}\frac{a^4}{l^4} - \frac{1107}{1024}\frac{a^5}{l^5} \right),
\end{eqnarray}
\begin{equation}\label{eq:sd2}
\omega = \frac{U}{a}\left(\frac{3}{4}\frac{a^2}{l^2} - \frac{9}{16}\frac{a^3}{l^3} + \frac{27}{64}\frac{a^4}{l^4} - \frac{177}{256}\frac{a^5}{l^5} + \frac{819}{1024}\frac{a^6}{l^6}\right),
\end{equation}
where $\eta$ is the absolute fluid viscosity.

\begin{table}[t]
\caption{\label{tab:sd} Table of $F/U$ and $\omega/U$ demonstrating good agreement between numerics (num) and theory (th) for sedimenting spheres, particularly for a large value of the ratio of sphere separation to radius, $l/a$.}
\begin{tabular}{c|cc|cc}
 $l/a$   & $F/U$(num) & $F/U$(th) & $\omega/U$(num) & $\omega/U$(th) \\ \hline
 10    & $1.3946988$ & $1.3946984$ & $0.006973076$ & $0.006973573$\\
 4     & $1.2551892$ & $1.2545428$ & $0.03906680$ & $0.03925395$\\
 3     & $1.1840827$ & $1.1806098$ & $0.06455729$ & $0.06596017$\\
 5/2   & $1.1308965$ & $1.1208450$ & $0.08549728$ & $0.09099600$\\
 25/12 & $1.0756197$ & $1.0450674$ & $0.09772490$ & $0.12515166$
\end{tabular}
\end{table}

 We simulated sedimenting spheres using $m = 0.25$, $I = (0.1, 0.1, 0.1)$, $l = 10$, 4, 3, 5/2, and 25/12, and time steps $h = 0.001$, 0.01, 0.02, 0.03, and 0.04 in  reduced units.  As expected for a 2-nd order numerical integrator, we observed linear dependence of the computed results on $h^2$.  A linear fit of $F/U$ and $\omega/U$ against $h^2$ was used to extrapolate the results to $h=0$, and these results are compared to the predictions of Eqs.~(\ref{eq:sd1}) and (\ref{eq:sd2}) in Table~\ref{tab:sd}. We see excellent agreement for large $l/a$, with the discrepancy increasing with decreasing $l/a$ due to the truncation of both the theoretical expression and the multipole expansion approximation in HYDROLIB.\\[3ex]

\noindent
{\bf Section S2: Lennard-Jones spheres in periodic boundary conditions}\\[1ex]
Having tested the implementation of HI in the noiseless limit, we now establish that the method converges to the correct Gibbsian distribution for a non-trivial model.
We underline that the experiment of this section only serves as a test that the proposed Langevin-type equations and numerical integrator sample accurately from the Gibbs distribution, as intended.  If sampling from the canonical ensemble of rigid bodies is the purpose  of the simulation, one can use the numerical integrators Langevin~A or Langevin~C from Ref.~\onlinecite{Davidchack15jcp_s}.

Here we perform simulation of $N$ interacting spheres in a cubic simulation box of size $L$ with periodic boundary conditions (PBC). In addition to the HI evaluated in HYDROLIB with enabled PBC option, the spheres interact with one another via a pairwise smoothly truncated Lennard-Jones (LJ) potential
\begin{equation} \label{eq:lj}
  u(x) = \left\{\begin{array}{ll} u_\mathrm{LJ}(x)\,, & x \leq x_m\,, \\
  u_\mathrm{LJ}(x)\phi(z(x))\,, & x_m < x < x_c\,, \\
  0\,, & x_c \leq x\, ,
  \end{array}\right.
\end{equation}
where $u_\mathrm{LJ}(x) = 4\epsilon\left[(\sigma/x)^{12} - (\sigma/x)^6\right]$, $\phi(z) = 1 - 10z^3 + 15z^4 - 6z^5$, $z(x) = (x^2 - x_m^2)/(x_c^2 - x_m^2)$.  The potential $u(x)$ is twice continuously differentiable. Here, $x$ is the distance between interaction sites; the interaction site of sphere $i$ is offset from the sphere centre-of-mass $r^i$ by a vector $d$ in the body frame, in order to induce rotational forces.  We set the size of the box $L$ sufficiently large so that $x_c \leq L/2$ and the minimum image convention applies.  As such, the total potential energy of the spheres is
\begin{equation} \label{eq:poten}
  U(\mathbf{r},\mathbf{q}) = \sum_{i=2}^N\sum_{j=1}^{i-1} u(|r^i + A\tran(q^i)d - r^j - A\tran(q^j)d|),
\end{equation}
where $q^i$ are the quaternion coordinates of sphere $i$.

\begin{figure}[t]
\begin{center}
  \includegraphics[width=8.2cm]{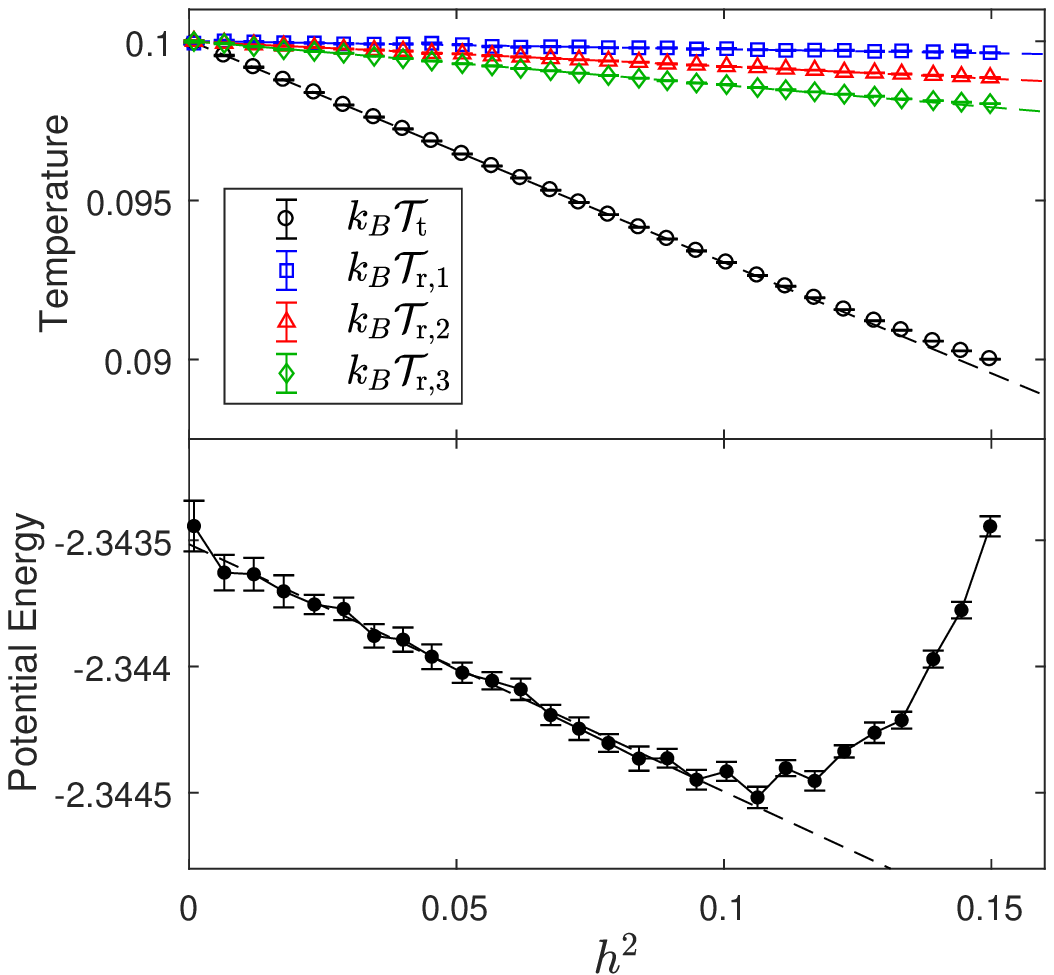}
\end{center}
\caption{\label{fig:t01} Spheres in PBC with LJ and hydrodynamic interactions. $N = 8$, $k_B T$ = 0.1.  Dashed lines indicate straight line weighted least-squares approximation
of the results for $h^2 < 0.1$. 
}
\end{figure}

\begin{figure}[t]
\begin{center}
  \includegraphics[width=8.2cm]{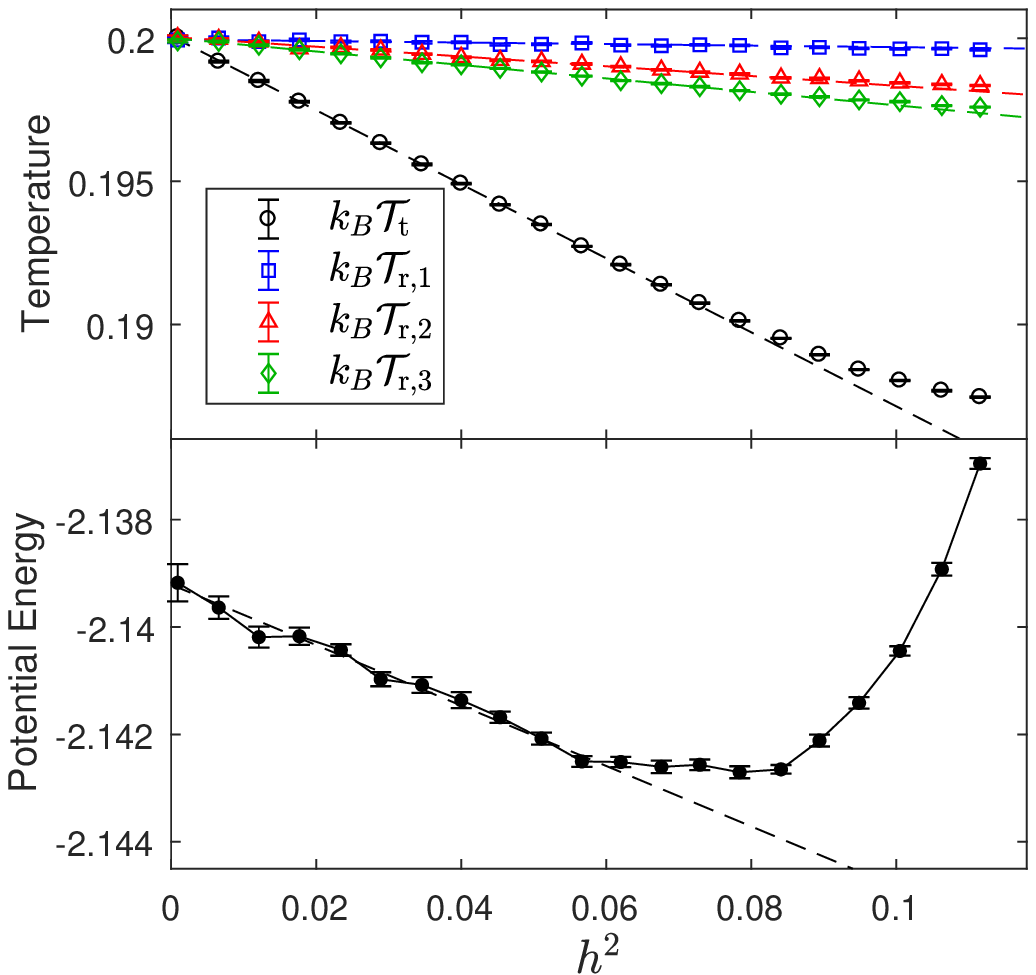}
\end{center}
\caption{\label{fig:t02} Spheres in PBC with LJ and hydrodynamic interactions. Here $N = 8$ and $k_B T$ = 0.2. Dashed lines indicate straight line weighted least-squares approximation of the results for $h^2 < 0.06$.}
\end{figure}

We augment the reduced units of the HYDROLIB package (the radius of the spheres is set to 1, the viscosity of the surrounding fluid $\eta$ is set to $1/(4\pi)$) with an energy scale by setting $\epsilon = 1$ in the Lennard-Jones potential.  The values of all other parameters and measured results are reported in these reduced units.

The numerical experiments were carried out with the following parameters: $N = 8$, $L = 15.0$, $\sigma = 2.6$.  The mass of the spheres is $m = 5.0$, the principal moments of inertia are $I = (3.0, 2.0, 1.5)$, and $d = (0.2, 0.15, 0)\tran$.  The LJ cut-off radius is $x_c = 2.5\,\sigma$, and $x_m = 0.9 x_c$.

Simulations were carried out at temperatures $\kb T = 0.1$ and 0.2, and a range of time steps $h$.  After 2000 equilibration steps, the measurements were taken over $2\times 10^5$ steps.  40 independent runs were performed at each $h$.  We measure temperature from kinetic energy of the spheres, separately for translational and rotational degrees of freedom (separately for each $l = 1$, 2, 3):
\begin{equation}\label{eq:ttra}
  \langle \mathcal{T}_\mathrm{t}\rangle_h  =
  \frac{\langle \mathbf{p}^{\mathsf{T}}\mathbf{p}\rangle_h }{3Nmk_B},
\end{equation}
\begin{equation}\label{eq:trot}
  \langle \mathcal{T}_{\mathrm{r},l}\rangle_h  =
  \frac{2\left\langle\sum_{j=1}^{N}V_{l}(q^{j},\pi ^{j})\right\rangle_h}{N k_B},
\end{equation}
as well as potential energy per sphere
\begin{equation}\label{eq:pot}
  \langle \mathcal{U} \rangle_h = \frac{1}{N} \langle U \rangle_h.
\end{equation}
The results are shown in Figures~\ref{fig:t01} and \ref{fig:t02}, where we see convergence of both translational and rotational temperatures to the thermostat parameter when $h \rightarrow 0$.  We observe linear dependence on $h^2$ for all measured quantities, as expected for a weak 2nd-order numerical integrator\cite{MT1_z}:
\begin{equation}\label{eq:error-h}
  \langle A \rangle_h = A_0 + E_A h^2 + \mathcal{O}(h^{3}),
\end{equation}
Estimated values of $A_0$ and $E_A$ for the measured quantities are shown in Table~\ref{tab:lsfit}.

\begin{table}[tb]
\caption{\label{tab:lsfit} Table of coefficients of least-square fit to the data at $h^2 < 0.1$ for $k_B T = 0.1$ and $h^2 < 0.06$ for $k_B T = 0.2$.}
\begin{tabular}{c|cc|cc}
 &\multicolumn{2}{c}{$k_B T = 0.1$} & \multicolumn{2}{c}{$k_B T = 0.2$}\\
 $A$ & $A_0$ & $E_A$ & $A_0$ & $E_A$\\ \hline
 $k_B \mathcal{T}_\mathrm{t}$     & $0.100012(14)$ & $-0.06972(11)$ & $0.20001(2)$ & $-0.1293(5)$\\
 $k_B \mathcal{T}_{\mathrm{r},1}$ & $0.100010(18)$ & $-0.00257(15)$ & $0.19998(3)$ & $-0.0027(7)$\\
 $k_B \mathcal{T}_{\mathrm{r},2}$ & $0.099992(14)$ & $-0.00778(12)$ & $0.20002(2)$ & $-0.0171(6)$\\
 $k_B \mathcal{T}_{\mathrm{r},3}$ & $0.100011(12)$ & $-0.01388(11)$ & $0.19999(3)$ & $-0.0235(5)$\\
 $\mathcal{U}$                & $-2.34352(3)$& $-0.0098(4)$ & $-2.13921(14)$ & $-0.056(3)$
\end{tabular}
\end{table}

\newpage

\noindent {\bf Section S3: Videos of circling spheres driven by external force and torque}
\begin{table}[h]\caption{List of videos of the system evolution.}
\begin{tabular}{cccl}
 $f$~~   & $\tau$~~ & $T$    & \hspace{10ex} file \\ \hline
 $1.0$~~   & $0$~~ & $0$    & \hspace{2ex}\href{https://www.dropbox.com/s/cgbs1g7dri2w5zk/ring_f1.0tau0T0.mp4?dl=0}{\tt ring\_f1.0tau0T0.mp4}\footnotemark[1]\\
 $1.2$~~   & $-4.0$~~ & $0$    & \hspace{2ex}\href{https://www.dropbox.com/s/amnenj18shzy02f/ring_f1.2tau-4.0T0.mp4?dl=0}{\tt ring\_f1.2tau-4.0T0.mp4}\footnotemark[2]\\
 $1.2$~~   & $-3.0$~~ & $0$    & \hspace{2ex}\href{https://www.dropbox.com/s/xrziqd7fs0ms892/ring_f1.2tau-3.0T0.mp4?dl=0}{\tt ring\_f1.2tau-3.0T0.mp4}\footnotemark[3]\\
 $0.4$~~   & $-4.0$~~ & $0$    & \hspace{2ex}\href{https://www.dropbox.com/s/331d57dl1hwmcn9/ring_f0.4tau-4.0T0.mp4?dl=0}{\tt ring\_f0.4tau-4.0T0.mp4}\footnotemark[4]\\
 $0$~~   & $-5.0$~~ & $0$    & \hspace{2ex}\href{https://www.dropbox.com/s/b6g8rpo0ggldgq6/ring_f0tau-5.0aT0.mp4?dl=0}{\tt ring\_f0tau-5.0aT0.mp4}\\
 $1.0$~~ & $0$~~    & $0.01$ & \hspace{2ex}\href{https://www.dropbox.com/s/v1e7mlalkaivrnt/ring_f1.0tau0T0.01.mp4?dl=0}{\tt ring\_f1.0tau0T0.01.mp4}\\
 $1.2$~~ & $-4.0$~~ & $0.01$ & \hspace{2ex}\href{https://www.dropbox.com/s/n0r2po44qh6dw5y/ring_f1.2tau-4.0T0.01.mp4?dl=0}{\tt ring\_f1.2tau-4.0T0.01.mp4}
\end{tabular}
\footnotetext[1]{See Fig.~7 in the main text.}
\footnotetext[2]{See Fig.~8(a) in the main text.}
\footnotetext[3]{See Fig.~8(b) in the main text.}
\footnotetext[4]{See Fig.~8(c) in the main text.}
\end{table}
\vspace{5ex}

\noindent {\bf Section S4: Implementation details of the numerical integrator}\\[1ex]
The numerical integrator (main text, Eq. (22)) was implemented in Fortran 90, using the HYDROLIB package \cite{hydrolib_s} to compute the friction matrix $\xi(\mathbf{r})$ and EXPOKIT \cite{expokit} subroutine {\tt dgpadm} to evaluate matrix exponents.  LAPACK subroutine {\tt dpotrf} was used to compute Cholesky factorisation. Below we provide the implementation details. The Ziggurat random number generator \cite{ziggurat,zigurl} was used for generating the Gaussian distribution.

HYDROLIB defines the number of particles variable {\tt \_NP\_} and global arrays for particle center-of-mass coordinates {\tt c(0:2,1:\_NP\_)}, linear and angular velocities {\tt v(1:6*\_NP\_)}, forces and torques {\tt f(1:6*\_NP\_)}, and the friction matrix {\tt fr(1:6*\_NP\_,1:6*\_NP\_)}.  Specifically, {\tt c(0:2,i)} contains components of $r^i$, {\tt v(6*i-5:6*i-3)} contains $p^i/m^i$, and {\tt f(6*i-5:6*i-3)} contains $f^i$.  The angular velocity and torque components in {\tt v} and {\tt f} are not used.  Instead, we define global variables {\tt qq(0:3,1:\_NP\_)}, {\tt qp(0:3,1:\_NP\_)}, and {\tt qf(0:3,1:\_NP\_)} containing $q^i$, $\pi^i$, and $F^i$, respectively, $i = 1,\ldots,n$.  Particles masses $m^i$ and principal moments of inertia $I^i = (I^i_1,I^i_2,I^i_3)\tran$, $i = 1,\ldots,n$, are stored in arrays {\tt mass(1:\_NP\_)} and {\tt inert(1:3,1:\_NP\_)}, respectively.  The structure of array {\tt fr} is described in the HYDROLIB guide.\\[1ex]

{\tt
\noindent SUBROUTINE OneStep\\
!\;one step of numerical integrator (main text, 22)\\
{\ind}hh = h/2\\
{\ind}do i = 1, \_NP\_\\
{\indd}i1 = 6*i-5;  i2 = i1+2\\
{\indd}v(i1:i2) = v(i1:i2) + hh*f(i1:i2)/mass(i)\\
{\indd}qp(:,i) = qp(:,i) + hh*qf(:,i)\\
{\indd}c(:,i) = c(:,i) + hh*v(i1:i2)\\
{\ind}end do\\
{\ind}CALL FreeRotorMinus(hh)\\
{\ind}CALL OUStep(h)\\
{\ind}do i = 1, \_NP\_\\
{\indd}i1 = 6*i-5;  i2 = i1+2\\
{\indd}c(:,i) = c(:,i) + hh*v(i1:i2)\\
{\ind}end do\\
{\ind}CALL FreeRotorPlus(hh)\\
{\ind}CALL ComputeForces\\
{\ind}do i = 1, \_NP\_\\
{\indd}i1 = 6*i-5;  i2 = i1+2\\
{\indd}v(i1:i2) = v(i1:i2) + hh*f(i1:i2)/mass(i)\\
{\indd}qp(:,i) = qp(:,i) + hh*qf(:,i)\\
{\ind}end do\\
END SUBROUTINE OneStep\\

\noindent SUBROUTINE FreeRotorMinus(dt)\\
!\;compute map $\Psi_{t}^{-}$ in Eq.\,(main text, 24)\\
{\ind}do i = 1, \_NP\_\\
{\indd}CALL Rotate(3, dt, i)\\
{\indd}CALL Rotate(2, dt, i)\\
{\indd}CALL Rotate(1, dt, i)\\
{\ind}end do\\
END SUBROUTINE FreeRotorMinus\\

\noindent SUBROUTINE FreeRotorPlus(dt)\\
!\;compute map $\Psi_{t}^{+}$ in Eq.\,(main text, 24)\\
{\ind}do i = 1, \_NP\_\\
{\indd}CALL Rotate(1, dt, i)\\
{\indd}CALL Rotate(2, dt, i)\\
{\indd}CALL Rotate(3, dt, i)\\
{\ind}end do\\
END SUBROUTINE FreeRotorPlus\\

\noindent SUBROUTINE Rotate(l, dt, i)\\
!\;compute map in Eq.\,({main text, 23})\\
{\ind}sq = Slq(l, qq(:,i))\\
{\ind}sp = Slq(l, qp(:,i))\\
{\ind}zdt = dt*dot\_product(qp(:,i),sq)/inert(l,i)/4\\
{\ind}qq(:,i) = cos(zdt)*qq(:,i) + sin(zdt)*sq\\
{\ind}qp(:,i) = cos(zdt)*qp(:,i) + sin(zdt)*sp\\
END SUBROUTINE Rotate\\

\noindent FUNCTION Slq(l, q)\\
{\ind}if (l == 1) then\\
{\indd}Slq(0:3) = [-q(1), q(0), q(3), -q(2)]\\
{\ind}else if (l == 2) then\\
{\indd}Slq(0:3) = [-q(2), -q(3), q(0), q(1)]\\
{\ind}else !\;l == 3\\
{\indd}Slq(0:3) = [-q(3), q(2), -q(1), q(0)]\\
{\ind}end if\\
END FUNCTION Slq\\

\noindent SUBROUTINE OUStep(h)\\
!\;compute Ornstein-Unlenbeck step
!\;HYDROLIB call:~compute $\xi(\mathbf{r})$, output in fr\\
{\ind}CALL Eval\\
!\;compute $\tilde{\xi}(\mathbf{r},\mathbf{q})$ in Eq.~(main text, 17), output in etxi\\
{\ind}do j = 1, \_NP\_\\
!\;compute $A\tran(q^j)\hat{D}^j \hat{S}\tran(q^j)$, output in ads\\
{\indd}ads = transpose(HatSq(j))\\
{\indd}ads(1,:)\;= ads(1,:)/inert(1,j)\\
{\indd}ads(2,:)\;= ads(2,:)/inert(2,j)\\
{\indd}ads(3,:)\;= ads(3,:)/inert(3,j)\\
{\indd}ads = matmul(transpose(RotMat(j),ads)\\
{\indd}kj = 6*j-5;  mj = 7*j-6\\
{\indd}do i = 1, \_NP\_\\
{\inddd}ki = 6*i-5;  mi = 7*i-6\\
{\inddd}etxi(mi:mi+2,mj:mj+2) = fr(ki:ki+2,kj:kj+2)/mass(j) !\;tt\\
{\inddd}etxi(mi:mi+2,mj+3:mj+6) = matmul(fr(ki:ki+2,kj+3:kj+5),ads)/2 !\;tr\\
{\inddd}etxi(mi+3:mi+6,mj:mj+2) = matmul(CheckSq(i),\\ fr(ki+3:ki+5,kj:kj+2))*2/mass(j) !\;rt\\
{\inddd}etxi(mi+3:mi+6,mj+3:mj+6) = matmul(CheckSq(i),\\ matmul(fr(ki+3:ki+5,kj+3:kj+5),ads)) !\;rr\\
{\indd}end do\\
{\ind}end do\\
!\;compute $\mathrm{e}^{-h\tilde{\xi}(\mathbf{r},\mathbf{q})}$, output in etxi\\
{\ind}CALL ExpMat(-h, 7*\_NP\_, etxi, etxi)\\
!\;compute $K\xi$, output in cc\\
{\ind}do i = 1, \_NP\_\\
!\;compute $A\tran(q^{i})\hat{D}^{i}A(q^{i})$, output in ada\\
{\indd}aa = RotMat(i) !\;$A(q^i)$\\
{\indd}do l = 1, 3\\
{\inddd}ada(l,:)\,= aa(l,:)/inert(l,i)\\
{\indd}end do\\
{\indd}ada = matmul(transpose(aa),ada)\\
{\indd}ki = 6*i-5\\
{\indd}do j = 1, \_NP\_\\
{\inddd}kj = 6*j-5\\
{\inddd}cc(ki:ki+2,kj:kj+5) = fr(ki:ki+2,kj:kj+5)/mass(i) !\;tt, tr\\
{\inddd}cc(ki+3:ki+5,kj:kj+2) = matmul(ada, fr(ki+3:ki+5,kj:kj+2)) !\;rt\\
{\inddd}cc(ki+3:ki+5,kj+3:kj+5) = matmul(ada, fr(ki+3:ki+5,kj+3:kj+5)) !\;rr\\
{\indd}end do\\
{\ind}end do\\
!\;compute $\mathrm{e}^{-2hK\xi}$, output in cc\\
{\ind}CALL ExpMat(-2*h, 6*\_NP\_, cc, cc)\\
!\;compute $-K^{-1}\left[\mathrm{e}^{-2hK \xi} - \bm{1}_{6n}\right]$, output in cc\\
{\ind}do ki = 1, 6*\_NP\_\\
{\indd}cc(ki,ki) = cc(ki,ki) - 1\\
{\ind}end do\\
{\ind}do i = 1, \_NP\_\\
!\;compute $-A^{\mathsf{T}}(q^{i})\left[\hat{D}^{i}\right]^{-1}A(q^{i})$, output in ada\\
{\indd}aa = RotMat(i) !\;$A(q^i)$\\
{\indd}do l = 1, 3\\
{\inddd}ada(l,:)\,= -inert(l,i)*aa(l,:)\\
{\indd}end do\\
{\indd}ada = matmul(transpose(aa), ada))\\
{\indd}ki = 6*i-5\\
{\indd}do j = 1, \_NP\_\\
{\inddd}kj = 6*j-5\\
{\inddd}cc(ki:ki+2,kj:kj+5) = -mass(i)*cc(it:tr+2,jt:jr+2) !\;tt, tr\\
{\inddd}cc(ki+3:ki+5,kj:kj+2) = matmul(ada,cc(ki+3:ki+5,kj:kj+2)) !\;rt\\
{\inddd}cc(ki+3:ki+5,kj+3:kj+5) = matmul(ada,cc(ki+3:ki+5,kj+3:kj+5)) !\;rr\\
{\indd}end do\\
{\ind}end do\\
!\;compute $\tilde{\sigma}$ by Cholesky factorization, output in cc\\
{\ind}CALL dpotrf('L', 6*\_NP\_, cc, 6*\_NP\_, info)\\
{\ind}do i = 1, 6*\_NP\_-1\\
{\indd}cc(i,i+1:6*\_NP\_) = 0\\
{\ind}end do\\
!\;compute $G_1\tilde{\sigma}$, output in sigma\\
{\ind}sigma = 0\\
{\ind}do i = 1, \_NP\_\\
{\indd}cs = 2*CheckSq(i) !\;$2\check{S}(q^i)$\\
{\indd}ki = 6*i-5; mi = 7*i-6\\
{\indd}do j = 1, i\\
{\inddd}kj = 6*(j-1)+1\\
{\inddd}sigma(mi:mi+2,kj:kj+5) = cc(ki:ki+2,kj:kj+5) !\;tt, tr\\
{\inddd}sigma(mi+3:mi+6,kj:kj+2) = matmul(cs,cc(ki+3:ki+5,kj:kj+2)) !\;rt\\
{\inddd}sigma(mi+3:mi+6,kj+3:kj+5) = matmul(cs,cc(ki+3:ki+5,kj+3:kj+5)) !\;rr\\
{\indd}end do\\
{\ind}end do\\
!\;define $Y = (\mathbf{r}\tran,\mathbf{q}\tran)\tran$, output in yy\\
{\ind}do i = 1, \_NP\_\\
{\indd}ki = 6*i-5; mi = 7*i-6\\
{\indd}yy(mi:mi+2) = mass(i)*v(ki:ki+2)\\
{\indd}yy(mi+3:mi+6) = qp(:,i)\\
{\ind}end do\\
!\;compute $\mathrm{e}^{-h\tilde{\xi}}Y$, output in yy\\
{\ind}yy = matmul(etxi, yy)\\
!\;generate $\chi$ with i.i.d.~$\mathcal{N}(0,1)$ components\\
{\ind}CALL RandNiid(6*\_NP\_, chi)\\
!\;add $\sigma(h;\mathbf{r},\mathbf{q})\chi$\\
{\ind}yy = yy + matmul(sigma, sqrt(tempr)*chi)\\
!\;copy from yy back to v and qp\\
{\ind}do i = 1, \_NP\_\\
{\indd}ki = 6*i-5; mi = 7*i-6\\
{\indd}v(ki:ki+2) = Y(mi:mi+2)/mass(i)\\
{\indd}qp(:,i) = Y(mi+3:mi+6)\\
{\ind}end do\\
END SUBROUTINE OUStep\\

\noindent SUBROUTINE ExpMat(t, n, hh, ethh)\\
!\;compute exp(t*hh(1:n,1:n)), output in ethh\\
{\ind}integer, parameter ::\;ideg = 6\\
{\ind}real ::\;wsp(4*n*n+ideg+1), ipiv(n)\\
{\ind}lwsp = 4*n*n+ideg+1\\
!\;EXPOKIT subroutine\\
{\ind}CALL dgpadm(ideg,n,t,hh,n,wsp,lwsp,ipiv,iexph,\\ ns,iflag)\\
ethh = reshape(wsp(iexph:iexph+n*n-1), [n, n])\\
END SUBROUTINE ExpMat\\

\noindent SUBROUTINE RandNiid(n, rnd)\\
{\ind}do i = 1, n\\
!\;call function from ziggurat.f90\cite{zigurl}\\
{\indd}rnd(i) = r4\_nor(seed, kn, fn, wn)\\
{\ind}end do\\
END SUBROUTINE RandNiid\\

\noindent FUNCTION RotMat(i)\\
{\ind}k = 0\\
{\ind}do l = 0, 3\\
{\indd}do m = l, 3\\
{\inddd}k = k + 1\\
{\inddd}p(k) = 2*qq(l,i)*qq(m,i)\\
{\indd}end do\\
{\ind}end do\\
{\ind}RotMat(1,1:3) = [p(1)+p(5)-1, p(6)+p(4), p(7)-p(3)]\\
{\ind}RotMat(2,1:3) = [p(6)+p(4), p(1)+p(8)-1, p(9)+p(2)]\\
{\ind}RotMat(3,1:3) = [p(7)+p(3), p(9)-p(2), p(1)+p(10)-1]\\
END FUNCTION RotMat\\

\noindent FUNCTION HatSq(i)\\
{\ind}HatSq(:,1) = [-qq(1,i),  qq(0,i),  qq(3,i), -qq(2,i)]\\
{\ind}HatSq(:,2) = [-qq(2,i), -qq(3,i),  qq(0,i),  qq(1,i)]\\
{\ind}HatSq(:,3) = [-qq(3,i),  qq(2,i), -qq(1,i),  qq(0,i)]\\
END FUNCTION HatSq(i)\\

\noindent FUNCTION CheckSq(i)\\
{\ind}S(:,1) = [-qq(1,i),  qq(0,i), -qq(3,i),  qq(2,i)]\\
{\ind}S(:,2) = [-qq(2,i),  qq(3,i),  qq(0,i), -qq(1,i)]\\
{\ind}S(:,3) = [-qq(3,i), -qq(2,i),  qq(1,i),  qq(0,i)]\\
END FUNCTION CheckSq(i)\\
}

%

\end{document}